\documentclass{article}
\usepackage{ijcai23}

\usepackage{times}
\usepackage{soul}
\usepackage{url}
\usepackage[hidelinks]{hyperref}
\usepackage[utf8]{inputenc}
\usepackage[small]{caption}
\usepackage{graphicx}
\usepackage{amsmath}
\usepackage{amsthm}
\usepackage{booktabs}
\usepackage{algorithm}
\usepackage{algorithmic}
\usepackage[switch]{lineno}
\usepackage{comment}
\usepackage{multirow}
\usepackage{amssymb}
\usepackage{type1cm}
\usepackage{stfloats}
\usepackage{color}

\newcommand{\argmin}{\mathop{\rm arg~min}\limits}
\renewcommand{\Vec}[1]{\textrm{\boldmath $#1$}} 

\title{Learning to Speak from Text: Zero-Shot Multilingual Text-to-Speech with Unsupervised Text Pretraining}

\author{
Takaaki Saeki$^1$
\and
Soumi Maiti$^2$
\and
Xinjian Li$^2$
\and
Shinji Watanabe$^2$
\and\\
Shinnosuke Takamichi$^1$
\And
Hiroshi Saruwatari$^1$
\affiliations
$^1$The University of Tokyo, Japan\\
$^2$Carnegie Mellon University, USA
\emails
takaaki\_saeki@ipc.i.u-tokyo.ac.jp,
\{smaiti, swatanab\}@andrew.cmu.edu
}

\begin{document}

\maketitle

\begin{abstract}
While neural text-to-speech (TTS) has achieved human-like natural synthetic speech, multilingual TTS systems are limited to resource-rich languages due to the need for paired text and studio-quality audio data.
This paper proposes a method for zero-shot multilingual TTS using text-only data for the target language.
The use of text-only data allows the development of TTS systems for low-resource languages for which only textual resources are available, making TTS accessible to thousands of languages.
Inspired by the strong cross-lingual transferability of multilingual language models, our framework first performs masked language model pretraining with multilingual text-only data.
Then we train this model with a paired data in a supervised manner, while freezing a language-aware embedding layer.
This allows inference even for languages not included in the paired data but present in the text-only data.
Evaluation results demonstrate highly intelligible zero-shot TTS with a character error rate of less than 12\% for an unseen language.
\end{abstract}

\section{Introduction}

Recent advances in end-to-end neural text-to-speech synthesis (TTS)~\cite{li2019neural,kim2021conditional} have yielded significant improvements in naturalness and speech quality.
However, the data-intensive nature and the requirement of paired text and studio-quality audio data have limited multilingual TTS systems to resource-rich languages, which are small portions of the more than 6,000 languages in the world~\cite{gordon2005ethnologue}.
To address the limitation, current research in multilingual TTS aims not only to exploit resource-rich languages~\cite{zen2012statistical,li2016multi} but also to build models for low-resource languages~\cite{prakash2019building}.

\begin{figure}
    \centering
    \includegraphics[width=0.76\linewidth, clip]{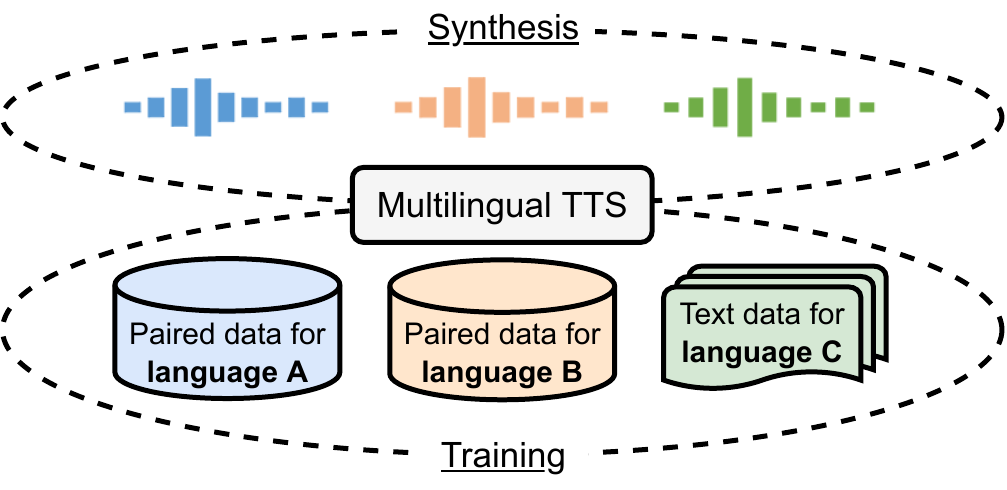}
    \caption{Our concept. We aim to build TTS model on languages for which only text data is available, to support low-resource languages.}
    \label{fig:overview}
\end{figure}

Previous work has addressed low-resource TTS by using untranscribed speech data with vector-quantized variational autoencoder (VQ-VAE)~\cite{zhang2020unsupervised} or automatic speech recognition (ASR) models~\cite{ni2022unsupervised}. Another study~\cite{saeki2022virtuoso} has built a massively multilingual TTS model jointly using paired TTS, paired ASR, unpaired speech, and unpaired text data.
However, these approaches still rely on speech data for the target languages and face the challenge of data collection, when audio recordings for these languages are hard to obtain.
In this study, we focus on the use of a text-only data for multilingual TTS as shown in Fig.~\ref{fig:overview}.
Previous research~\cite{wu2019beto,pires2019multilingual} has shown the strong cross-lingual transferability of multilingual language models such as multilingual BERT~\cite{devlin2019bert} in natural language processing (NLP) tasks.
By leveraging multilingual pretraining, the model can generalize to other languages, even if it has never seen the target data in those languages.
Our work applies the framework of multilingual masked language model (MLM) pretraining to TTS, with the goal of achieving zero-shot cross-lingual transfer of pronunciation and prosody.
Zero-shot TTS using text data enables the development of TTS systems for languages where only textual resources are available, which potentially opens up TTS to thousands of languages~\cite{ebrahimi2021adapt,li2022asr2k}.

\begin{figure}
    \centering
    \includegraphics[width=0.94\linewidth, clip]{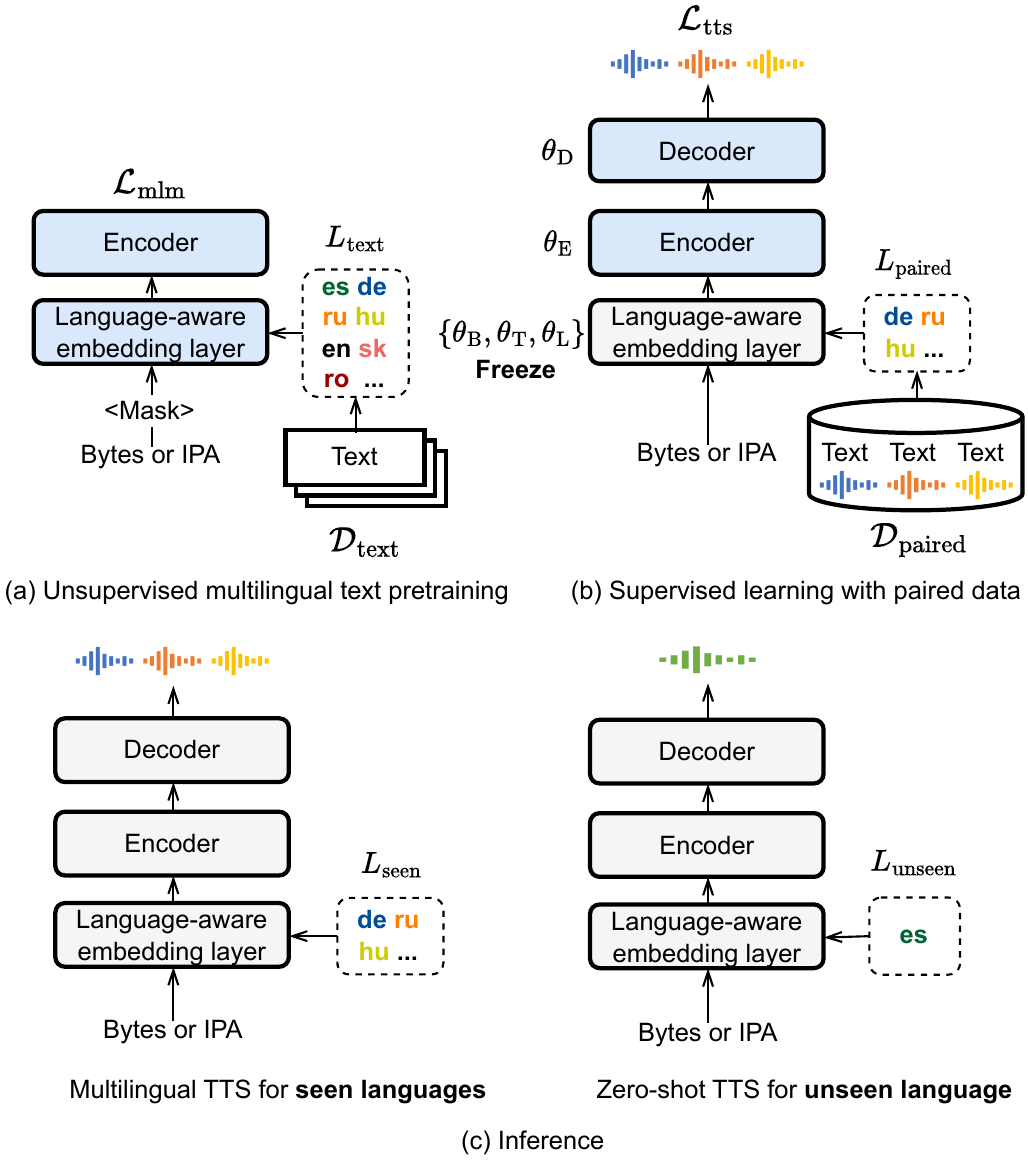}
    \caption{Proposed framework. (a) We perform MLM pretraining on multilingual text data and then (b) train TTS model on paired data with frozen language-aware embedding layer. (c) Zero-shot TTS is performed with language IDs that are not included in paired data.}
    \label{fig:method}
\end{figure}

In this paper, we propose a multilingual TTS framework that leverages unsupervised text pretraining.
Fig.~\ref{fig:method} illustrates the proposed framework.
We use a typical end-to-end TTS architecture consisting of token embedding, encoder, and decoder.
Our model also has a language-aware embedding layer, which includes the token embedding layer, a language embedding layer, and a bottleneck layer.
As shown in Fig.~\ref{fig:method}(a), we first pretrain the language-aware embedding layer and the encoder of the TTS model with multilingual text data.
We then fine-tune the encoder and decoder of the TTS model with paired data, while the language-aware embedding layer is frozen, as illustrated in Fig.~\ref{fig:method}(b).
This allows zero-shot TTS for a language not included in the paired data but present in the text data, as shown on the right in Fig.~\ref{fig:method}(c).

Our contributions are as follows.
1) We propose a zero-shot multilingual TTS framework that achieves highly intelligible TTS for an unseen language, resulting in a character error rate of less than 12\%.
2) Our method also improves TTS for seen languages, resulting in byte-based models without grapheme-to-phoneme (G2P) modules that outperform the phoneme-based baselines.
3) Our ablation studies provide additional insights, including the effectiveness of the frozen language-aware embedding layer.
The experiments were conducted on public datasets and the implementation is available\footnote{\url{https://github.com/Takaaki-Saeki/zm-text-tts}}.
We encourage readers to listen to our audio samples\footnote{\url{https://takaaki-saeki.github.io/zm-tts-text_demo}}.

\section{Method}
Our model has a typical neural TTS model architecture consisting of token embedding, encoder, and decoder. 
First, we use MLM pretraining with multilingual text data to learn cross-lingual representations.
Then we perform supervised learning with paired data to learn the mapping from linguistic features to speech features.
The model performs inference even for languages that are not present in the paired data.

\subsection{Unsupervised Multilingual Text Pretraining}\label{sec:method-pretrain}
Fig.~\ref{fig:method}(a) illustrates the unsupervised pretraining method.
It uses multilingual text data consisting of languages that are not included in the paired data.
Let $X = (x_{n} \in V | n = 1, \cdots, N)$ denote the input text token sequence of length $N$, where $V$ denotes a vocabulary constructed for pretraining.
We define $\mathcal{D}_{\mathrm{text}}$ as the text dataset.
Let $L_{\mathrm{text}}$ denote the set of language IDs included in $\mathcal{D}_{\mathrm{text}}$.
First, the masked token sequence $X^{\mathrm{m}}$ and a language ID $l_{\mathrm{text}} \in L_{\mathrm{text}}$ are fed to the model.
Let the token embedding sequence and language embedding be $Z^{\mathrm{m}} = (\Vec{z}^{\mathrm{m}}_{n} \in \mathbb{R}^{d} | n = 1, \cdots, N)$ and $\Vec{e}_{l} \in \mathbb{R} ^ {d}$, respectively.
The embedding layers output $Z^{\mathrm{m}}$ and $\Vec{e}_{l}$ as:
\begin{equation}\label{eq:embed-t-l}
\begin{split}
Z^{\mathrm{m}} = \text{Embed}(X^{\mathrm{m}}; \theta_{\mathrm{T}}), \qquad
\Vec{e}_{l} = \text{Embed}(l_{\mathrm{text}}; \theta_{\mathrm{L}}),
\end{split}
\end{equation}
where $\theta_{\mathrm{T}}$ and $\theta_{\mathrm{L}}$ denote the model parameters of the token embedding and language embedding layers, respectively.
Then the token and language embeddings obtained in Eq.~\eqref{eq:embed-t-l} are added and fed to a bottleneck layer to project them into a hidden input vector.
Let $H_{\mathrm{in}} = (\Vec{h}_{\mathrm{in}, n} \in \mathbb{R}^{d} | n = 1, \cdots, N)$ and $H_{\mathrm{out}} = (\Vec{h}_{\mathrm{out}, n} \in \mathbb{R}^{d} | n = 1, \cdots, N)$ denote hidden vectors in the encoder input and output, respectively.
Then the conditional probability $p(X|X_{-\Pi})$ is computed as:
\begin{align}
H_\mathrm{in} &= \text{Bottleneck}(Z^{\mathrm{m}} + \Vec{e}^{l}; \theta_{\mathrm{B}}), \label{eq:encoder-mlm} \\
H_\mathrm{out} &= \text{Encoder}(H_\mathrm{in}; \theta_{\mathrm{E}}), \label{eq:encoder-mlm-1} \\
p(X|X_{-\Pi}) &= \text{Softmax}(\text{PredictionNet}(H_{\mathrm{out}}; \theta_{\mathrm{P}})), \label{eq:encoder-mlm-2}
\end{align}
where $\theta_{\mathrm{B}}$, $\theta_{\mathrm{E}}$, $\theta_{\mathrm{P}}$ denote the model parameters of the bottleneck layer, the encoder and a prediction network, respectively.
In Eq.~\eqref{eq:encoder-mlm-2}, $\text{Softmax}(\cdot)$ denotes a softmax function.
We define the network with the model parameters $\{\theta_{\mathrm{B}}, \theta_{\mathrm{T}}, \theta_{\mathrm{L}}\}$ as \textbf{language-aware embedding layer}, which jointly embeds the token sequence $X$ and the language ID $l_{\mathrm{text}}$ as in Eq.~\eqref{eq:embed-t-l} and \eqref{eq:encoder-mlm}.
Let $\Pi = (\pi_{k} \in \mathbb{N} | k = 1, \cdots, K)$ be the indexes of the masked tokens of length $K$.
With the probability computed in Eq.~\eqref{eq:encoder-mlm-2}, the training objective can be defined as:
\begin{equation}\label{eq:mlm-loss}
\begin{split}
\mathcal{L}_{\mathrm{mlm}} &= \frac{1}{K} \sum_{k=1}^{K} \log p(x_{\pi_{k}} | X^{\mathrm{m}}), \\
\{ \hat{\theta}_{\mathrm{E}}, \hat{\theta}_{\mathrm{B}}, \hat{\theta}_{\mathrm{T}}, \hat{\theta}_{\mathrm{L}} \} &= \argmin_{\theta_{\mathrm{E}}, \theta_{\mathrm{B}}, \theta_{\mathrm{T}}, \theta_{\mathrm{L}}} \mathcal{L}_{\mathrm{mlm}}.
\end{split}
\end{equation}

We use UTF-8 bytes or International Phonetic Alphabet (IPA) symbols for the input token sequence $X$.
For each token type, the vocabulary $V$ is constructed from $\mathcal{D}_{\mathrm{text}}$, which includes a start/end of sentence token ([\texttt{SOS/EOS}]).
We extracted International IPA sequences using an open-source toolkit\footnote{\url{https://github.com/espeak-ng/espeak-ng}}.
To obtain the masked token $X^{\mathrm{m}}$, we use the same masking ratio and category as in the original BERT pre-training~\cite{devlin2019bert} for each token type.
Randomly, 12~\% of the tokens are replaced with the [\texttt{MASK}] token, and 1.5~\% of them are replaced with random tokens.
Also, 1.5~\% of the tokens are left unchanged and $\mathcal{L}_{\mathrm{mlm}}$ is computed as in Eq.~\eqref{eq:mlm-loss} for those 15~\% of tokens that have indices $\Pi$.

\subsection{Supervised Learning with Paired Data}\label{sec:method-supervised}
Fig.~\ref{fig:method}(b) illustrates the supervised learning of the TTS model with paired data.
We define the paired data and the set of language IDs as $\mathcal{D}_{\mathrm{paired}}$ and $L_{\mathrm{paired}}$, respectively.
Note that we assume $L_{\mathrm{paired}} \subset L_{\mathrm{text}}$.
Let $Y = (\Vec{y}_{t} \in \mathbb{R}^{D} | t=1, \cdots, T)$ denote the speech feature sequence with the length of $T$.
We first initialize the model parameters $ \{\theta_{\mathrm{E}}, \theta_{\mathrm{B}}, \theta_{\mathrm{T}}, \theta_{\mathrm{L}} \}$ with those obtained in the pretraining described in \S~\ref{sec:method-pretrain}.
Let $\theta_{\mathrm{D}}$ denote the model parameter of the decoder.
The speech features are predicted with teacher forcing as:
\begin{align}
H_{\mathrm{out}} &= \text{Encoder}(\text{Bottleneck}(Z + \Vec{e}^{l})), \label{eq:supervised-enc} \\
\hat{Y} &= \text{Decoder}(H_{\mathrm{out}}, Y; \theta_{\mathrm{D}}), \label{eq:supervised-dec}
\end{align}
where $Z$ is the unmasked token embedding sequence.
Note that the unmasked token sequence is used in Eq.~\eqref{eq:supervised-enc}, while the masked token sequence is used in Eq.~\eqref{eq:encoder-mlm}
Let $\mathcal{L}_{\mathrm{tts}}(\hat{Y}, Y)$ denote the training objective of the TTS model.
Then we consider two types of schemes. 

\paragraph{Updating language-aware embedding layer}
We only freeze the parameter of the language embedding layer $\theta_{\mathrm{L}}$ while updating the rest of the parameters.
Therefore the trainable model parameters can be written as
\begin{equation}\label{eq:supervised-update}
\{ \hat{\theta}_{\mathrm{D}}, \hat{\theta}_{\mathrm{E}}, \hat{\theta}_{\mathrm{B}}, \hat{\theta}_{\mathrm{T}} \} = \argmin_{\theta_{\mathrm{D}}, \theta_{\mathrm{E}}, \theta_{\mathrm{B}}, \theta_{\mathrm{T}}} \mathcal{L}_{\mathrm{tts}}(\hat{Y}, Y).
\end{equation}
Previous work has confirmed that multilingual BERT has high cross-lingual transferability for various NLP tasks~\cite{wu2019beto}.
This scheme corresponds to a simple fine-tuning of BERT~\cite{wu2019beto}, which updates all the parameters during training for the downstream tasks\footnote{We freeze the language embedding layer to address the mismatch between language embedding of seen and unseen languages.}.

\paragraph{Freezing language-aware embedding layer}
We freeze the bottleneck layer and the token embedding layer along with the language embedding, updating the encoder and decoder.
The training process can be written as
\begin{equation}\label{eq:supervised-freeze}
\{ \hat{\theta}_{\mathrm{D}}, \hat{\theta}_{\mathrm{E}} \} = \argmin_{\theta_{\mathrm{D}}, \theta_{\mathrm{E}}} \mathcal{L}_{\mathrm{tts}}(\hat{Y}, Y).
\end{equation}
In contrast to the scheme represented in Eq.~\eqref{eq:supervised-update}, the scheme in Eq.~\eqref{eq:supervised-freeze} preserves the parameters of the language-aware embedding layer to facilitate cross-lingual transfer. In the evaluation, we use the scheme formulated in Eq.~\eqref{eq:supervised-freeze}, except for the ablation study in \S~\ref{sec:eval-ablation}.

\subsection{Inference}\label{sec:method-inference}
Let $L_{\mathrm{syn}}$ denote the set of language IDs used for inference.
The text token sequence $X$ and the language ID $l_{\mathrm{syn}} \in L_{\mathrm{syn}}$ are fed to the model as in Eq.~\eqref{eq:embed-t-l}, and the encoder output is predicted as in Eq.~\eqref{eq:supervised-enc}.
Unlike Eq.~\eqref{eq:supervised-dec}, the speech features are predicted as:
\begin{equation}
\hat{Y} = \text{Decoder}(H_{\mathrm{out}}; \theta_{\mathrm{D}}).
\end{equation}
The output waveform is obtained by feeding the predicted features $\hat{Y}$ to a pretrained neural vocoder.

Fig.~\ref{fig:method}(c) illustrates the inference process.
The left and right sides of the figure show the typical multilingual TTS and our zero-shot TTS.
Previous work~\cite{li2019bytes} has typically assumed \textit{seen} languages, and the inference is performed with the language IDs $L_{\mathrm{seen}} \subset L_{\mathrm{paired}}$.
However, it is challenging to perform TTS for \textit{unseen} languages $L_{\mathrm{unseen}} \cap L_{\mathrm{paired}} = \emptyset$.
While other work~\cite{saeki2022virtuoso} has built a massively multilingual TTS model that even achieves zero-shot TTS from ASR data, it uses paired data for the target languages.
Our work attempts to only use the linguistic knowledge to improve the zero-shot TTS.
Thus, the inference process is written as $L_{\mathrm{unseen}}^{\prime} \cap L_{\mathrm{paired}} = \emptyset$ and $L_{\mathrm{unseen}}^{\prime} \subset L_{\mathrm{text}}$.
In the evaluation, we denote the inference with $L_{\mathrm{unseen}} $ and $L_{\mathrm{unseen}}^{\prime}$ as \textit{Fully zero-shot TTS} and \textit{Text-seen zero-shot TTS}, respectively.
\textit{Fully zero-shot TTS} performs zero-shot TTS without pretraining as in the IPA-based previous method~\cite{staib2020phonological}, which is the baseline method in our evaluations.

\subsection{Model Architecture}\label{sec:method-architecture}
Our model is an autoregressive TTS model based on Transformer TTS~\cite{li2019neural}, which has also been used in the previous work on byte-based multilingual TTS~\cite{he2021multilingual}.
During the supervised learning described in \S~\ref{sec:method-supervised} and inference described in \S~\ref{fig:method}, we use x-vector~\cite{snyder2018x} for the speaker embedding and add it to the encoder output through a projection layer.
During supervised learning, we use the average x-vectors computed from the training data.
For evaluation purposes, we perform zero-shot synthesis with the average x-vector from the test data of the target language and feed it to the model.
Note that we also conduct the evaluation with x-vectors from seen languages.

For the bottleneck layer with $\theta_{\mathrm{B}}$, we use a residual network consisting of Layer Normalization~\cite{ba2016layer}, down projection, ReLU~\cite{nair2010rectified}, and up projection with the residual connection, which is used in previous work on language adaptation~\cite{bapna2019simple}.

\section{Experimental Evaluations}\label{sec:eval}

\subsection{Experimental Setting}\label{sec:eval-setting}

\subsubsection{Dataset}\label{sec:eval-setting-dataset}
We carried out all the evaluations with publicly available datasets.
Table~\ref{tab:data} shows the sizes of the data for each language.
For the unsupervised text pretraining described in \S~\ref{sec:method-pretrain}, we used transcripts from VoxPopuli~\cite{wang2021voxpopuli}, M-AILABS~\cite{mailabs}, and CSS10~\cite{park2019css10}, resulting in a total of about 2.8~GB of spoken text across 19 languages.
We used CSS10 for the supervised learning described in \S~\ref{sec:method-supervised}, and we selected seven European languages as the seen languages, with Spanish as the unseen language.
The paired data consisted of one speaker per language.
It should be noted that Spanish is not actually a low-resource language, but we chose to use it for evaluation purposes in order to 1) compare our zero-shot TTS methods with the oracle methods using the paired data for the target language and 2) ensure a sufficient number of evaluators for the subjective evaluation.
We used 5 and 100 utterances as dev and test sets, respectively, with the remaining data used for training.

\begin{table}[tb]
\centering
\scalebox{0.74}{
\begin{tabular}{lcccc}
\toprule
\multicolumn{1}{l|}{\multirow{2}{*}{Languages}} & \multicolumn{1}{l|}{\multirow{2}{*}{Code}} & \multicolumn{1}{l|}{\multirow{2}{*}{Text-only data}} & \multicolumn{2}{c}{Paired data}              \\
\multicolumn{1}{l|}{}                           & \multicolumn{1}{l|}{}                      & \multicolumn{1}{l|}{}                                & \multicolumn{1}{c|}{Text}   & Audio          \\ \midrule
\multicolumn{5}{l}{\textit{Seen languages for evaluation} $L_{\mathrm{seen}}$}                                                                                                                                         \\ \midrule
\multicolumn{1}{l|}{German}                     & \multicolumn{1}{c|}{de}                    & \multicolumn{1}{c|}{359MB}                           & \multicolumn{1}{c|}{0.73MB} & 16.13h         \\
\multicolumn{1}{l|}{French}                     & \multicolumn{1}{c|}{fr}                    & \multicolumn{1}{c|}{372MB}                           & \multicolumn{1}{c|}{0.94MB} & 19.15h         \\
\multicolumn{1}{l|}{Dutch}                      & \multicolumn{1}{c|}{nl}                    & \multicolumn{1}{c|}{336MB}                           & \multicolumn{1}{c|}{0.75MB} & 14.10h         \\
\multicolumn{1}{l|}{Finnish}                    & \multicolumn{1}{c|}{fi}                    & \multicolumn{1}{c|}{308MB}                           & \multicolumn{1}{c|}{0.47MB} & 21.36h         \\
\multicolumn{1}{l|}{Hungarian}                  & \multicolumn{1}{c|}{hu}                    & \multicolumn{1}{c|}{104MB}                           & \multicolumn{1}{c|}{0.51MB} & 10.53h         \\
\multicolumn{1}{l|}{Russian}                    & \multicolumn{1}{c|}{ru}                    & \multicolumn{1}{c|}{4.9MB}                           & \multicolumn{1}{c|}{1.5MB}  & 10.00h         \\
\multicolumn{1}{l|}{Greek}                      & \multicolumn{1}{c|}{el}                    & \multicolumn{1}{c|}{0.39MB}                          & \multicolumn{1}{c|}{0.39MB} & 4.13h          \\ \midrule
\multicolumn{5}{l}{\textit{Unseen language for evaluation} $L_{\mathrm{unseen}}$}                                                                                                                                        \\ \midrule
\multicolumn{1}{l|}{Spanish}                    & \multicolumn{1}{c|}{es}                    & \multicolumn{1}{c|}{345MB}                           & \multicolumn{1}{c|}{0.0MB (1.2MB)} & 0.00h (23.81h) \\ \midrule
\multicolumn{5}{l}{Languages not included in CSS10}                                                                                                                                                \\ \midrule
\multicolumn{1}{l|}{English}                    & \multicolumn{1}{c|}{en}                    & \multicolumn{1}{c|}{338MB}                           & \multicolumn{1}{c|}{}    &          \\
\multicolumn{1}{l|}{Estonian}                   & \multicolumn{1}{c|}{et}                    & \multicolumn{1}{c|}{87MB}                            & \multicolumn{1}{c|}{}    &           \\
\multicolumn{1}{l|}{Croatian}                   & \multicolumn{1}{c|}{hr}                    & \multicolumn{1}{c|}{2.0MB}                           & \multicolumn{1}{c|}{}    &          \\
\multicolumn{1}{l|}{Italian}                    & \multicolumn{1}{c|}{it}                    & \multicolumn{1}{c|}{334MB}                           & \multicolumn{1}{c|}{}    &          \\
\multicolumn{1}{l|}{Lithuanian}                 & \multicolumn{1}{c|}{lt}                    & \multicolumn{1}{c|}{89MB}                            & \multicolumn{1}{c|}{}    &           \\
\multicolumn{1}{l|}{Polish}                     & \multicolumn{1}{c|}{pl}                    & \multicolumn{1}{c|}{102MB}                           & \multicolumn{1}{c|}{}    &          \\
\multicolumn{1}{l|}{Romanian}                   & \multicolumn{1}{c|}{ro}                    & \multicolumn{1}{c|}{67MB}                            & \multicolumn{1}{c|}{}    &          \\
\multicolumn{1}{l|}{Slovak}                     & \multicolumn{1}{c|}{sk}                    & \multicolumn{1}{c|}{94MB}                            & \multicolumn{1}{c|}{}    &          \\
\multicolumn{1}{l|}{Slovenian}                  & \multicolumn{1}{c|}{sl}                    & \multicolumn{1}{c|}{81MB}                            & \multicolumn{1}{c|}{}    &          \\ \bottomrule
\end{tabular}
}
\caption{Amount of text-only and paired data for each language. Parentheses indicate amount of original data in CSS10.}
\label{tab:data}
\end{table}

\subsubsection{Training Details}\label{sec:eval-setting-training}
The sampling rate was set to 16~kHz.
An 80-dimension of mel filter bank, 1024 samples of FFT length, and 256 samples of frame shit were used for speech analysis.
For the pretraining described in \S~\ref{sec:method-pretrain}, we trained the model for 1.2M iterations using the Noam optimizer~\cite{vaswani2017attention} with the learning rate and warm-up step set to 1.0 and 10000, respectively.
For the TTS model described in \S~\ref{sec:method-architecture}, we used a 6-block Transformer encoder~\cite{vaswani2017attention} and a 6-block Transformer decoder, with a postnet consisting of five convolutional layers with a kernel size of five. The attention dimension and the number of attention heads were set to 512 and 8, respectively.
For the bottleneck layer described in \S~\ref{sec:method-architecture}, we set the hidden dimension after the down projection to 256.
The $\text{PredictionNet}$ in Eq.~\eqref{eq:encoder-mlm-2} consisted of a linear layer, a GELU activation function~\cite{hendrycks2016gaussian}, Layer Normalization, and a linear layer with the hidden dimension of 512. 
We also used guided attention loss~\cite{tachibana2018efficiently} to improve the training efficiency.
For the supervised learning described in \S~\ref{sec:method-supervised}, we trained the models for 2.47M~iterations (200~epochs).
The Noam optimizer was used with the warm-up step of 50000.
For the neural vocoder, we trained HiFi-GAN~\cite{kong2020hifi} for 2M~iterations with LibriTTS~\cite{zen2019libritts}, VCTK~\cite{veaux2017cstr}, and CSS10.
For the x-vector described in \S~\ref{sec:method-architecture}, we used a model trained on VoxCeleb1 and VoxCeleb2~\cite{nagrani2017voxceleb} published in SpeechBrain~\cite{ravanelli2021speechbrain}.
We used ESPnet2-TTS~\cite{watanabe2018espnet,hayashi2021espnet2} for the implementation.

\subsubsection{Baselines}\label{sec:eval-setting-baselines}
We developed baseline models without the pretraining.

\paragraph{Seen language}
\textit{Monolingual: }
We trained a model for each language independently.
Our preliminary study found that Transformer TTS was unstable\footnote{The original paper~\cite{li2019neural} also reports the instability.} and could not synthesize intelligible speech in the monolingual condition due to the lack of training data. Therefore, we used Tacotron2~\cite{shen2018natural} only for the monolingual models, as in the original paper of the dataset~\cite{park2019css10}.
\textit{Multilingual w/o LIDs: }
We trained a multilingual Transformer TTS model using the paired data shown in Table~\ref{tab:data} without language IDs (LIDs).
\textit{Multilingual w/ LIDs: }
We trained a multilingual Transformer TTS model with the paired data of the unseen language. It also used the language IDs.

\paragraph{Unseen language} We compared \textit{Fully zero-shot TTS} and \textit{Text-seen zero-shot TTS} defined in \S~\ref{sec:method-inference}.
In \textit{Oracle}, we used the \textit{Monolingual} and \textit{Multilingual w/ LIDs}, which used the paired data of the unseen language.
In \textit{Fully zero-shot TTS}, we used \textit{Multilingual w/o LIDs} to synthesize speech from text tokens in the unseen language. This method corresponds to the conventional multilingual TTS model using bytes~\cite{he2021multilingual} or IPA symbols~\cite{staib2020phonological}. 

\subsubsection{Evaluation Metrics}\label{sec:eval-setting-metrics}
To objectively measure the synthetic speech quality, we used mel cepstral distortion (MCD)~\cite{fukada92melcep} with the mel cepstrum dimension set to 25.
We also evaluated the intelligibility using CERs computed with a multilingual ASR model~\cite{radford2022robust}. We used a pretrained \textit{large} model that is publicly available\footnote{\url{https://github.com/openai/whisper}}.
To evaluate the naturalness, we carried out listening tests to calculate five-scale mean opinion scores (MOS) of synthesized speech for each method. Forty native speakers were recruited through Amazon Mechanical Turk~\cite{paolacci2010running} for each of the tests.
Furthermore, we leveraged a publicly available automatic MOS (AMOS) prediction model~\cite{saeki2022utmos} to evaluate the naturalness.
Note that the model was trained on English and Chinese datasets, but previous work~\cite{seki2022text} has reported that it also showed a correlation coefficient higher than 0.8 for another language (Japanese).

\begin{table*}[tb]
\centering
\scalebox{0.80}{
\begin{tabular}{lcccccccccccccc}
\toprule
\multicolumn{1}{l|}{\multirow{2}{*}{Method}}      & \multicolumn{2}{c|}{de}                            & \multicolumn{2}{c|}{fr}                            & \multicolumn{2}{c|}{ru}                               & \multicolumn{2}{c|}{fi}                            & \multicolumn{2}{c|}{hu}                            & \multicolumn{2}{c|}{nl}                             & \multicolumn{2}{c}{el}         \\
\multicolumn{1}{l|}{}                             & MCD           & \multicolumn{1}{c|}{CER}           & MCD           & \multicolumn{1}{c|}{CER}           & MCD           & \multicolumn{1}{c|}{CER}              & MCD           & \multicolumn{1}{c|}{CER}           & MCD           & \multicolumn{1}{c|}{CER}           & MCD           & \multicolumn{1}{c|}{CER}            & MCD           & CER            \\ \midrule
\multicolumn{1}{l|}{Natural}                      & -             & \multicolumn{1}{c|}{2.75}          & -             & \multicolumn{1}{c|}{4.52}          & -             & \multicolumn{1}{c|}{2.12}             & -             & \multicolumn{1}{c|}{4.73}          & -             & \multicolumn{1}{c|}{4.86}          & -             & \multicolumn{1}{c|}{6.22}           & -             & 7.14           \\ \midrule
\multicolumn{15}{l}{\textit{Baseline (Monolingual)}}                                                                                                                                                                                                                                                                                                                                                                  \\ \midrule
\multicolumn{1}{l|}{Bytes monolingual}            & 7.70          & \multicolumn{1}{c|}{8.61}          & 11.76         & \multicolumn{1}{c|}{91.82}         & 11.43         & \multicolumn{1}{c|}{\textgreater 100} & 8.33          & \multicolumn{1}{c|}{56.03}         & 10.22         & \multicolumn{1}{c|}{93.05}         & 7.49          & \multicolumn{1}{c|}{15.33}          & 10.20         & 85.98          \\
\multicolumn{1}{l|}{IPA monolingual}           & 7.38          & \multicolumn{1}{c|}{4.07}          & 8.96          & \multicolumn{1}{c|}{17.86}         & 11.89         & \multicolumn{1}{c|}{25.30}            & 7.23          & \multicolumn{1}{c|}{27.62}         & 7.59          & \multicolumn{1}{c|}{24.62}         & 7.80          & \multicolumn{1}{c|}{19.20}          & 8.16          & 21.79          \\ \midrule
\multicolumn{15}{l}{\textit{Baseline (Multilingual)}}                                                                                                                                                                                                                                                                                                                                                                 \\ \midrule
\multicolumn{1}{l|}{Bytes multilingual w/o LIDs}           & 7.68          & \multicolumn{1}{c|}{37.46}         & 8.71          & \multicolumn{1}{c|}{41.35}         & 9.38          & \multicolumn{1}{c|}{45.92}            & 6.26          & \multicolumn{1}{c|}{29.19}         & 6.48          & \multicolumn{1}{c|}{33.82}         & 8.46          & \multicolumn{1}{c|}{46.33}          & 7.64          & 36.24          \\
\multicolumn{1}{l|}{Bytes multilingual w/ LIDs}   & 6.51          & \multicolumn{1}{c|}{13.19}         & 10.84         & \multicolumn{1}{c|}{55.79}         & 12.89         & \multicolumn{1}{c|}{\textgreater 100} & 6.78          & \multicolumn{1}{c|}{27.22}         & 9.09          & \multicolumn{1}{c|}{42.97}         & 8.47          & \multicolumn{1}{c|}{39.37}          & 7.25          & 23.56          \\
\multicolumn{1}{l|}{IPA multilingual w/o LIDs} & 6.31          & \multicolumn{1}{c|}{10.64}         & 7.44          & \multicolumn{1}{c|}{20.86}         & 8.10          & \multicolumn{1}{c|}{35.32}            & 5.53          & \multicolumn{1}{c|}{19.56}         & 5.59          & \multicolumn{1}{c|}{14.03}         & 7.76          & \multicolumn{1}{c|}{34.49}          & 6.90          & 19.33          \\
\multicolumn{1}{l|}{IPA multilingual w/ LIDs}  & 6.16          & \multicolumn{1}{c|}{9.76}          & 6.88          & \multicolumn{1}{c|}{14.97}         & 7.63          & \multicolumn{1}{c|}{23.54}            & 5.17          & \multicolumn{1}{c|}{10.63}         & 5.28          & \multicolumn{1}{c|}{9.11}          & 6.95          & \multicolumn{1}{c|}{19.48}          & 6.90          & 16.97          \\ \midrule
\multicolumn{15}{l}{\textit{Proposed (Unsupervised text pretraining)}}                                                                                                                                                                                                                                                                                                                                                          \\ \midrule
\multicolumn{1}{l|}{Bytes multilingual}           & \textbf{5.65} & \multicolumn{1}{c|}{\textbf{3.79}} & \textbf{6.48} & \multicolumn{1}{c|}{\textbf{7.15}} & 7.38          & \multicolumn{1}{c|}{\textbf{10.62}}   & \textbf{4.99} & \multicolumn{1}{c|}{\textbf{5.28}} & \textbf{5.01} & \multicolumn{1}{c|}{\textbf{6.05}} & \textbf{6.52} & \multicolumn{1}{c|}{\textbf{13.74}} & 6.57          & 11.75          \\
\multicolumn{1}{l|}{IPA multilingual}          & 5.88          & \multicolumn{1}{c|}{5.52}          & 6.61          & \multicolumn{1}{c|}{7.72}          & \textbf{7.25} & \multicolumn{1}{c|}{15.85}            & 5.18          & \multicolumn{1}{c|}{8.62}          & 5.30          & \multicolumn{1}{c|}{7.37}          & 7.00          & \multicolumn{1}{c|}{14.42}          & \textbf{6.53} & \textbf{11.06} \\ \bottomrule
\end{tabular}
}
\caption{Evaluation results for \textit{seen} languages. Bold indicates best scores in baseline and proposed methods.}
\label{tab:comparison-seen-obj}
\end{table*}

\subsection{Evaluation Results on Seen Languages}\label{sec:eval-seen}
We evaluated our framework on the seen languages included in the paired data, as defined in \S~\ref{sec:method-inference}.
Table~\ref{tab:comparison-seen-obj} lists the results in MCD and CER.
Lower values are better for both metrics.
As we can see, the byte-based or IPA-based models with the proposed multilingual pretraining performed the best across all languages and metrics.
Among the baselines, byte-based monolingual and multilingual models tended to have higher MCD and CER than IPA-based models, and failed to synthesize intelligible speech in some languages.
For example, the baseline byte-based models showed the high CER values for French, which has a deep orthography, meaning that a single character has different pronunciations depending on the context.
We observed that our method improved the byte-based models and they outperformed the IPA-based baseline models for all the metrics and languages.
It is worth noting that the proposed byte-based models even outperformed the proposed IPA-based models except for el and ru.
These results suggest that our framework is effective in building a TTS model for languages without G2P modules.

\begin{table}[tb]
\centering
\scalebox{0.86}{
\begin{tabular}{lccc}
\toprule
\multicolumn{1}{l|}{\multirow{3}{*}{Method}}        & \multicolumn{3}{c}{es}                           \\
\multicolumn{1}{l|}{}                               & \multicolumn{2}{c}{es x-vector}    & fr x-vector \\
\multicolumn{1}{l|}{}                               & MCD   & \multicolumn{1}{c|}{CER}   & CER         \\ \midrule
\multicolumn{1}{l|}{Natural}                        & -     & \multicolumn{1}{c|}{2.71}  & 2.71        \\ \midrule
\multicolumn{4}{l}{\textit{Oracle}}                                                                    \\ \midrule
\multicolumn{1}{l|}{Bytes monolingual}              & 8.65  & \multicolumn{1}{c|}{10.70} & -           \\
\multicolumn{1}{l|}{IPA monolingual}           & 8.47  & \multicolumn{1}{c|}{5.28}  & -           \\
\multicolumn{1}{l|}{IPA multilingual}          & 6.20  & \multicolumn{1}{c|}{5.32}  & 6.99        \\ \midrule
\multicolumn{4}{l}{\textit{Baseline (Fully zero-shot TTS)}}                                                               \\ \midrule
\multicolumn{1}{l|}{Bytes multilingual}             & 11.22 & \multicolumn{1}{c|}{64.07} & 66.45       \\
\multicolumn{1}{l|}{IPA multilingual}          & 10.75 & \multicolumn{1}{c|}{44.75} & 44.37       \\ \midrule
\multicolumn{4}{l}{\textit{Proposed (Text-seen zero-shot TTS)}}                                             \\ \midrule
\multicolumn{1}{l|}{Bytes multilingual}             & 9.05  & \multicolumn{1}{c|}{18.27} & 13.74       \\
\multicolumn{1}{l|}{IPA multilingual}          & 9.44  & \multicolumn{1}{c|}{11.69} & 13.33       \\ \bottomrule
\end{tabular}
}
\caption{Evaluation results for \textit{unseen} language.}
\label{tab:comparison-unseen-obj}
\end{table}

\begin{table*}[tb]
\centering
\scalebox{0.82}{
\begin{tabular}{l|cccccccc|cc|cc}
\toprule
\multirow{3}{*}{Method}  & \multicolumn{8}{c|}{Seen}                                                                                                         & \multicolumn{2}{c|}{Unseen} & \multicolumn{2}{c}{\multirow{2}{*}{Avg.}} \\
                         & \multicolumn{2}{c}{de}           & \multicolumn{2}{c}{fr}           & \multicolumn{2}{c}{ru}            & \multicolumn{2}{c|}{fi} & \multicolumn{2}{c|}{es}     & \multicolumn{2}{c}{}                     \\
                         & MCD  & \multicolumn{1}{c|}{CER}  & MCD  & \multicolumn{1}{c|}{CER}  & MCD  & \multicolumn{1}{c|}{CER}   & MCD        & CER        & MCD          & CER          & MCD                & CER                 \\ \midrule
Bytes multilingual       & 5.65 & \multicolumn{1}{c|}{3.79} & 6.48 & \multicolumn{1}{c|}{7.15} & 7.38 & \multicolumn{1}{c|}{10.62} & 4.99       & 5.28       & 9.05         & 18.27        & 6.46               & \textbf{9.58}                \\ \midrule
W/o bottleneck layer     & 6.06 & \multicolumn{1}{c|}{5.01} & 7.15 & \multicolumn{1}{c|}{9.09} & 7.71 & \multicolumn{1}{c|}{28.52} & 5.33       & 6.47       & 10.26        & 24.01        & 6.99               & 13.74               \\
W/o language ID          & 6.07 & \multicolumn{1}{c|}{5.09} & 7.09 & \multicolumn{1}{c|}{9.99} & 7.77 & \multicolumn{1}{c|}{22.58} & 5.23       & 6.99       & 10.45        & 32.70        & 6.96               & 14.06               \\
W/o initializing encoder & 5.59 & \multicolumn{1}{c|}{3.75} & 6.52 & \multicolumn{1}{c|}{9.31} & 7.12 & \multicolumn{1}{c|}{16.47} & 4.86       & 5.03       & 9.02         & 21.91        & \textbf{6.42}               & 11.85               \\
Updating language-aware embedding layer               & 6.05 & \multicolumn{1}{c|}{6.22} & 6.75 & \multicolumn{1}{c|}{6.93} & 7.46 & \multicolumn{1}{c|}{11.42} & 5.16       & 8.00       & 9.48         & 17.21        & 6.75               & 10.62               \\ \bottomrule
\end{tabular}
}
\caption{Ablation studies on training and model configurations. Bold indicates best metrics on average (Avg.).}
\label{tab:ablation}
\end{table*}

\subsection{Evaluation Results on Unseen Language}\label{sec:eval-unseen}

We evaluated our method on zero-shot TTS for the unseen language defined in \S~\ref{sec:method-inference}.
As described in \S~\ref{sec:method-architecture}, we first used the x-vector from the es speaker to compute the MCD.
Table~\ref{tab:comparison-unseen-obj} lists the results. 
The baseline models showed the CERs of over 40\% and MCDs of over 10.0.
However, our proposed text preraining improved the metrics, resulting in CERs of less than half for both byte and IPA-based methods.
Also, in contrast to the results for the seen languages, the IPA-based model outperformed the byte-based one in terms of CER.
Compared with the oracle case with the paired data of the unseen language, our proposed zero-shot TTS showed higher MCD and CER but achieved only 1\% difference in CER compared to the oracle byte-based monolingual model.
These results demonstrate the effectiveness of our method in achieving intelligible zero-shot TTS for the unseen language.

To investigate the case where the target speaker information is completely unavailable, we also used the x-vector from a seen language. We chose the fr speaker because es and fr are both categorized as Western Romance in Glottolog~\cite{hammarstrom2021glottolog}.
Table~\ref{tab:comparison-unseen-obj} lists the results.
Note that this case does not have the MCD results, since a different speaker than the ground-truth speech was used.
We can see that the unsupervised text pretraining also improved the zero-shot performance when using the x-vector from the fr speaker.
In the proposed byte-based model, the cross-lingual x-vector showed the lower CER.
This might result from that the es x-vector was not present in the training data whereas the fr x-vector was present in the training data.

\begin{table}[tb]
\centering
\scalebox{0.88}{
\begin{tabular}{lcccc}
\toprule
\multicolumn{1}{l|}{\multirow{2}{*}{Method}} & \multicolumn{2}{c|}{de}        & \multicolumn{2}{c}{hu} \\
\multicolumn{1}{l|}{}                        & MCD   & \multicolumn{1}{c|}{CER}   & MCD           & CER           \\ \midrule
\multicolumn{1}{l|}{Natural}                 & -     & \multicolumn{1}{c|}{2.75}  & -             & 2.12          \\ \midrule
\multicolumn{5}{l}{\textit{Oracle}}                                                                               \\ \midrule
\multicolumn{1}{l|}{IPA monolingual}      & 7.38  & \multicolumn{1}{c|}{4.07}  & 7.59          & 24.62             \\
\multicolumn{1}{l|}{IPA multilingual}     & 6.16  & \multicolumn{1}{c|}{9.76}  & 5.28          & 9.11          \\ \midrule
\multicolumn{5}{l}{\textit{Baseline (Fully zero-shot TTS)}}                                                                          \\ \midrule
\multicolumn{1}{l|}{IPA multilingual}     & 10.31 & \multicolumn{1}{c|}{38.75} & 9.93          & 52.62         \\ \midrule
\multicolumn{5}{l}{\textit{Proposed (Text-seen zero-shot TTS)}}                                                        \\ \midrule
\multicolumn{1}{l|}{Bytes multilingual}      & 10.00 & \multicolumn{1}{c|}{28.01} & 9.40          & 50.11         \\\bottomrule
\end{tabular}
}
\caption{Analysis on different \textit{unseen} languages.}
\label{tab:diff-unseen}
\end{table}

\begin{figure}
    \centering
    \includegraphics[width=0.90\linewidth, clip]{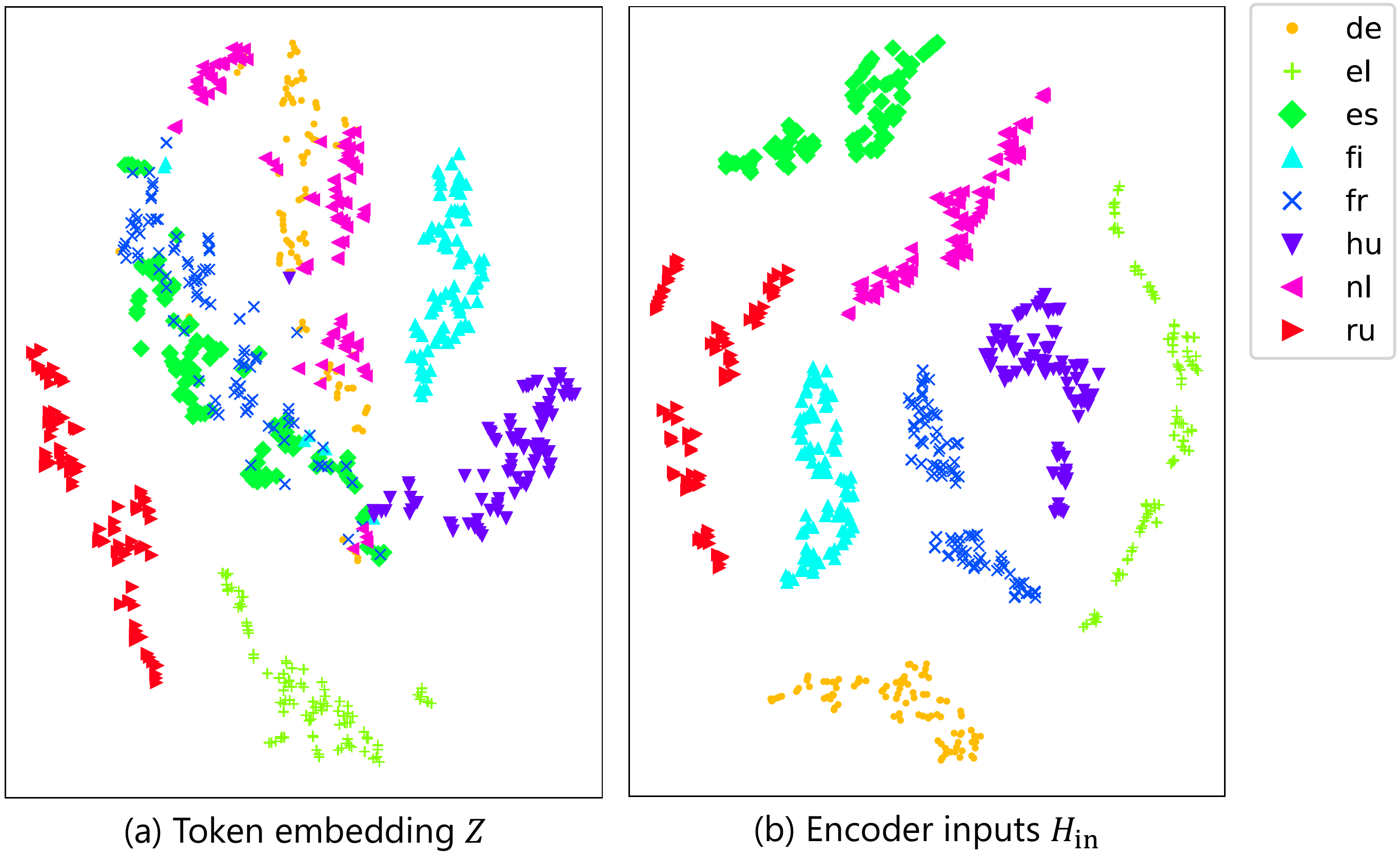}
    \caption{Visualization of token and language embedding. Pairs of similar languages (es--fr and de--nl) are overlapping in token embedding space, while output of bottleneck layer separates them.}
    \label{fig:visualize}
\end{figure}

\begin{figure*}
    \centering
    \includegraphics[width=0.90\linewidth, clip]{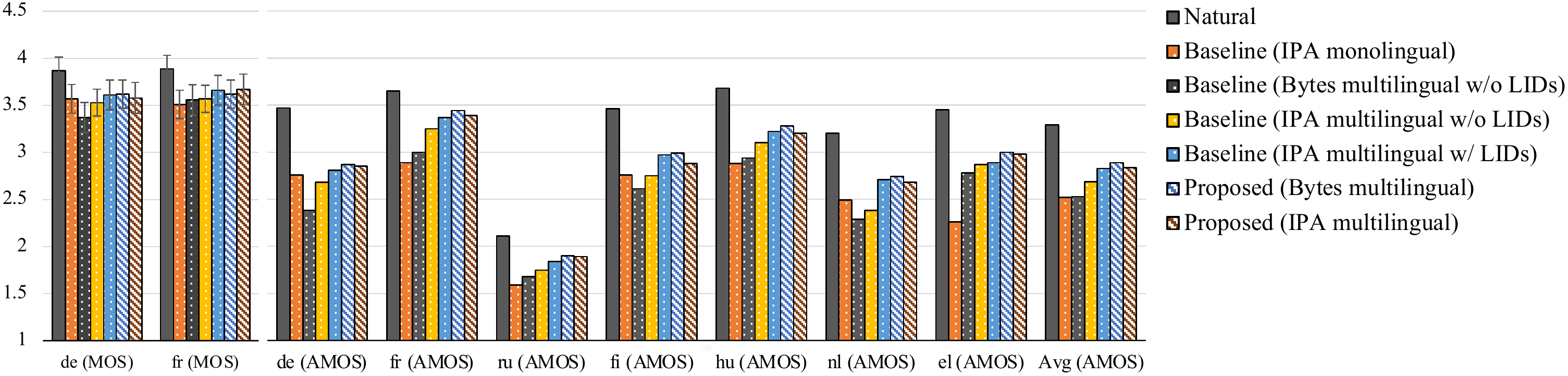}
    \caption{MOS and AMOS results for \textit{seen} languages. Error bars in MOS results represent 95\% confidence intervals.}
    \label{fig:mos-pmos-seen}
\end{figure*}

\subsection{Ablation Study}\label{sec:eval-ablation}

To further evaluate our method, we conducted several ablation studies.
Table~\ref{tab:ablation} lists the results.
\textit{Bytes multilingual} represents the byte-based proposed method in the evaluation of \S~\ref{sec:eval-seen} and \ref{sec:eval-unseen}. 
Note that it used the frozen language-aware embedding layer as formulated in Eq.~\eqref{eq:supervised-freeze}.
Some additional studies of our method are also presented in the Appendix.

In \textit{W/o bottleneck layer}, we excluded the bottleneck layer and simply added the token and language embedding to obtain the encoder input in Eq.~\eqref{eq:encoder-mlm}.
We found that removing the bottleneck layer led to a performance drop in all the languages and metrics, with an average increase of 0.53 in MCD and 4.16\% in CER.
The largest increase was observed in the unseen language, with an increase of 1.21 in MCD.
This suggests that the bottleneck layer, which projects the token and language embedding into the hidden input text representation with nonlinear dimensionality reduction, is effective in improving the generalization for zero-shot TTS.

We also evaluated the effect of including language IDs in the proposed method by comparing it with a version that excluded language IDs, referred to as \textit{W/o language ID}.
It corresponds to a simple multilingual BERT pretraining~\cite{wu2019beto} that uses only text tokens across different languages.
We observed that the use of language IDs led to an average improvement of 0.5 MCD and 4.48\% CER, indicating the effectiveness of our approach in using language IDs.

In \textit{W/o initializing encoder}, we did not initialize the encoder $\theta_{\mathrm{E}}$ before the supervised leaning described in \S~\ref{sec:method-supervised}. Instead, we only initialized the parameters $\theta_{\mathrm{T}}$, $\theta_{\mathrm{L}}$, and $\theta_{\mathrm{B}}$ with the parameters pretrained in \S~\ref{sec:method-pretrain}.
Through this evaluation, we investigated whether the performance gain with our method resulted from the initialization of the language-aware embedding layer or the encoder.
We observed that \textit{W/o initializing encoder} resulted in an improvement of 0.04 in MCD and only a 2.27\% increase in CER on average, suggesting that our method benefits more from the pretraining of the language-aware embedding layer than from the encoder.

In \textit{Updating language-aware embedding layer}, we updated the language-aware embedding layer during supervised learning, as formulated in Eq.~\eqref{eq:supervised-update}. We observed that freezing the language-aware embedding layer led to better performance for most languages and metrics, resulting in an average difference of 0.29 in MCD and 1.04\% in CER.

\begin{figure}
    \centering
    \includegraphics[width=0.98\linewidth, clip]{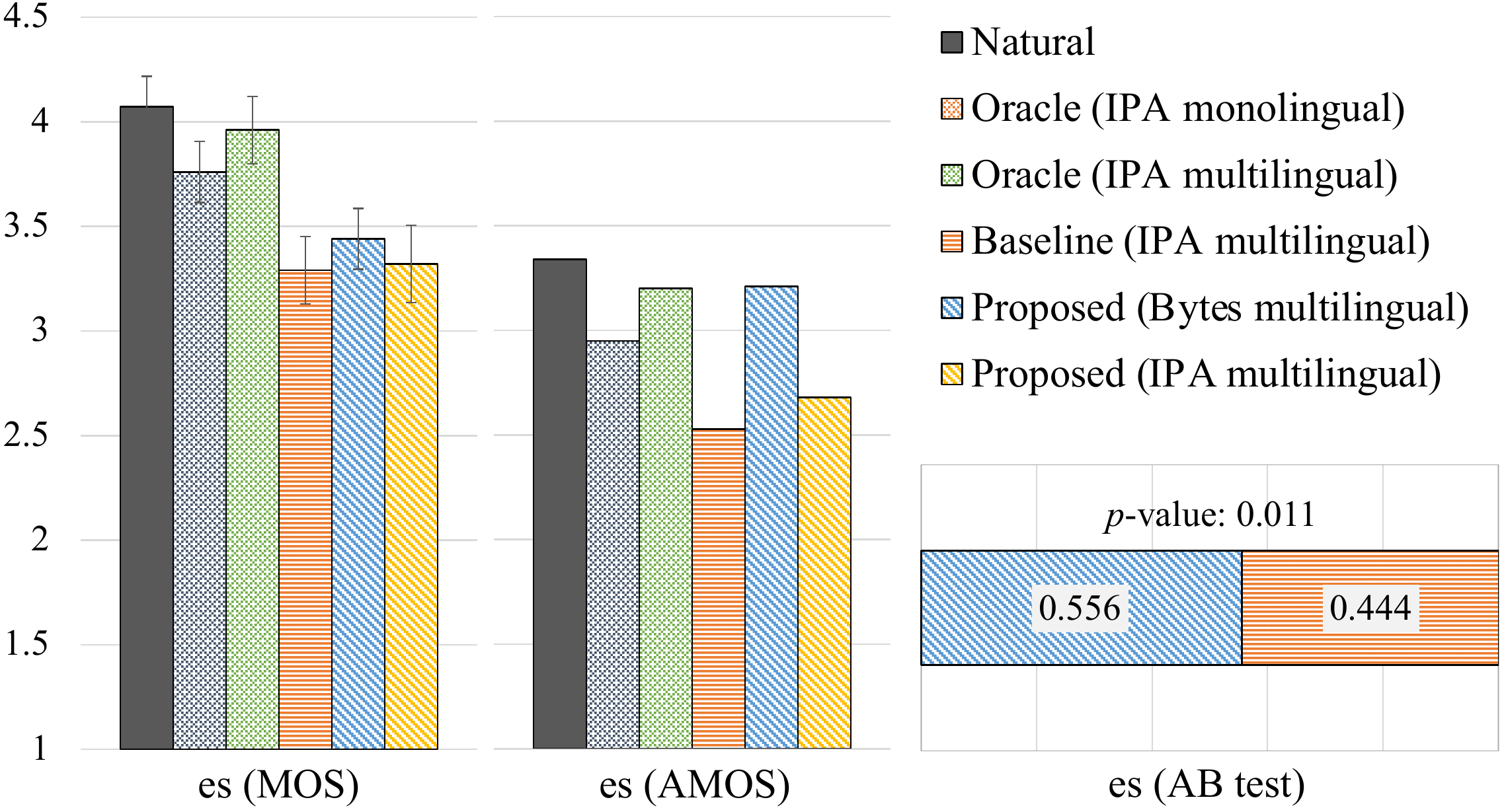}
    \caption{MOS, AMOS, and AB test results for \textit{unseen} language. Error bars in MOS results represent 95\% confidence intervals.}
    \label{fig:mos-pmos-unseen}
\end{figure}

\subsection{Dependency on Unseen Languages}\label{sec:eval-dependency}
We conducted evaluations on the zero-shot TTS for different unseen languages.
The eight European languages included in the paired data are composed of Indo-European and Uralic language families defined in Glottolog~\cite{hammarstrom2021glottolog}.
In this evaluation, we selected de and hu from each of the families.
During supervised learning in \S~\ref{sec:method-supervised}, we excluded the paired data for each of de and hu and instead included the paired data for es.
Table~\ref{tab:diff-unseen} lists the results.
We chose the IPA-based baseline method, which had shown better results in \S~\ref{sec:eval-unseen}. 
We observed that the pretraining improved the CER by around 10\% and MCD by around 0.3 for de.
However, the improvement in CER for hu was limited to 2\%, while the MCD was improved by around 0.5.
These results suggest that the performance of our zero-shot TTS is language dependent, as observed in previous work on cross-lingual transfer for NLP tasks~\cite{wu2019beto}.

Fig.~\ref{fig:visualize} visualize the token embedding $Z$ and encoder inputs $H_{\mathrm{in}}$ averaged on each utterance. We used a t-distributed stochastic neighbor embeddings (t-SNE)~\cite{van2008visualizing}.
We observed overlaps in the token embedding for (es, fr) and (de, nl), which are classified as Western Romance and West Germanic in Glottolog, respectively.
The encoder inputs are separated in the embedding space for each language.
The results in Table~\ref{tab:diff-unseen} and the visualization suggest that the cross-lingual transfer works better when similar languages sharing the token embedding space are present during supervised learning.
However, for languages with distinct token and language embeddings, the cross-lingual transferability might be limited.
We leave the further analysis on language dependencies as a topic for future research.

\subsection{Subjective Evaluations on Naturalness}\label{sec:eval-subjective}
We conducted evaluations on naturalness as described in \S~\ref{sec:eval-setting-metrics}.
Fig.~\ref{fig:mos-pmos-seen} shows the results for seen languages.
Note that we conducted the listening tests for de and fr.
For each language, either of the proposed methods showed the highest MOS, while we did not observe any significant difference between the proposed methods and the best baseline method, which was the IPA-based multilingual model with LIDs.
To further validate our results, we also evaluated the naturalness with an AMOS prediction model, as shown in Fig.~\ref{fig:mos-pmos-seen}.
We observed that the either of the proposed methods showed the highest scores in all the languages.
On average, the byte-based and IPA-based proposed models showed 2.89 and 2.84, respectively, while the best baseline method obtained 2.83\footnote{The AMOS tended to be lower than the MOS. While the MOS prediction model has a high correlation, it may produce errors in predicting absolute values, as reported in previous work~\cite{saeki2022utmos}. The relative relationships are more reliable in the AMOS.}.
Additionally, we observed that the byte-based proposed model often scored higher than the IPA-based proposed models, which is consistent with the results in Table~\ref{tab:comparison-seen-obj}.

Fig.~\ref{fig:mos-pmos-unseen} shows the results for unseen languages.
The oracle methods had the highest MOS of 3.76 and 3.96, and the baseline zero-shot method had the lowest MOS of 3.29.
The proposed methods outperformed the baseline method, and the byte- and IPA-based models had the MOS of 3.44 and 3.32, respectively. 
The AMOS results were consistent with the listening test results, with the proposed zero-shot TTS methods outperforming the baseline method.
In this evaluation, the proposed byte-based model scored 3.21 on the AMOS, while the oracle IPA-based model scored 3.20.
To further validate the results, we conducted a preference AB test on naturalness with 25~rators. As shown in Fig.~\ref{fig:mos-pmos-unseen}, our byte-based model significantly outperformed the baseline IPA-based model.

\section{Related Work}

\paragraph{Multilingual TTS}
While previous work on multilingual TTS has primarily focused on resource-rich languages~\cite{zen2012statistical,li2016multi}, there is growing interest in developing TTS models on low-resource languages.
Several studies have explored the input tokens shared across languages such as bytes~\cite{li2019bytes,he2021multilingual}, IPA symbols~\cite{gutkin2017uniform}, and articulatory features~\cite{lux2022language}, to transfer knowledge from resource-rich to low-resource languages.
Grapheme tokens can eliminate the per-language G2P knowledge, and previous work has built a byte-based TTS model for around 40 languages~\cite{he2021multilingual}.
There has been work using the phonological features derived from IPA to achieve the zero-shot TTS~\cite{staib2020phonological}.
Our framework achieves the zero-shot cross-lingual transfer with bytes by leveraging multilingual text pretraining.
There have been studies on using untranscribed speech data for low-resource scenarios by leveraging VQ-VAE~\cite{zhang2020unsupervised} or an ASR model~\cite{ren2019almost,ni2022unsupervised}.
Other work~\cite{saeki2022virtuoso} has trained a massively multilingual TTS using paired TTS, paired ASR, unpaired speech, and unpaired text data. While it also performs text-only training as in our work, it still uses the paired speech-text data of the target languages.
Our framework is simple and scalable, while pioneering a novel paradigm with the zero-shot TTS approach that relies only on text data.

\paragraph{Cross-lingual representation learning for NLP}
There have been studies on learning cross-lingual representations that can be applied to various NLP tasks in different languages~\cite{gouws2015bilbowa,ruder2019survey}.
Recent work has highlighted the strong cross-lingual transferability of multilingual BERT~\cite{devlin2019bert}, which has been observed to perform surprisingly well when transferred to other languages~\cite{wu2019beto,lample2019cross}.
Building on this, our work leverages multilingual MLM pretraining for TTS, which improves byte-based TTS models without G2P knowledge and achieves zero-shot TTS.

\paragraph{Language model pretraining for TTS}
Previous research has explored self-supervised text pretraining techniques for TTS.
BERT models have been used to extract contextual embeddings and enhance the prosody of TTS~\cite{hayashi2019pre,xu2021improving}. Other studies have used phonemes jointly with graphemes~\cite{jia2021png} or sub-phonemes~\cite{zhang2022mixed} as the inputs of the MLM pretraining.
Our work proposes multilingual MLM pretraining for TTS using text tokens shared across languages, rather than focusing on monolingual pretraining.

\section{Conclusions}

We presented a multilingual TTS framework that leverages unsupervised text pretraining.
Our framework achieved highly intelligible zero-shot TTS for an unseen language, resulting in a CER of less than 12\%.
It also improved the TTS for seen languages, with byte-based models without G2P modules outperforming the IPA-based baselines.
Our ablation studies provided additional insights, including the effectiveness of the frozen language embedding layer.

\paragraph{Limitations and future work}
Our proposed framework has limitations.
The performance gap remains between the oracle models and our zero-shot TTS models in terms of intelligibility, speech quality, and naturalness, as seen in the evaluation in \S~\ref{sec:eval-unseen} and \S~\ref{sec:eval-subjective}. Further studies are needed to improve our zero-shot TTS.
Our framework also has a limitation with language dependency, as the results in \S~\ref{sec:eval-dependency} suggest that this dependency is caused by the presence of similar languages during supervised learning.
Our future work will focus on studying this language dependency further and developing a method that performs better for various languages.

\section*{Acknowledgments}
Part of this work was supported by JSPS KAKENHI Grant Number 21H05054, 22H03639, and 22J12040.
This work used the Bridges system~\cite{nystrom2015bridges}, which is supported by NSF
award number ACI-1445606, at the Pittsburgh Supercomputing Center.
We would like to thank the research teams at Google through the internship of the first author for providing various insights on this topic.

\bibliographystyle{named}
\bibliography{ijcai23}

\begin{thebibliography}{}

\bibitem[\protect\citeauthoryear{Ba \bgroup \em et al.\egroup
  }{2016}]{ba2016layer}
J.~L. Ba, J.~R. Kiros, and G.~E Hinton.
\newblock Layer normalization.
\newblock {\em arXiv preprint arXiv:1607.06450}, 2016.

\bibitem[\protect\citeauthoryear{Ba{\~n}{\'o}n \bgroup \em et al.\egroup
  }{2020}]{banon2020paracrawl}
M.~Ba{\~n}{\'o}n, P.~Chen, B.~Haddow, K.~Heafield, H.~Hoang,
  M.~Espl{\`a}-Gomis, M.~L Forcada, A.~Kamran, F.~Kirefu, P.~Koehn, et~al.
\newblock Paracrawl: Web-scale acquisition of parallel corpora.
\newblock In {\em Proc. ACL}, pages 4555--4567, 2020.

\bibitem[\protect\citeauthoryear{Bapna and Firat}{2019}]{bapna2019simple}
A.~Bapna and O.~Firat.
\newblock Simple, scalable adaptation for neural machine translation.
\newblock In {\em Proc. EMNLP-IJCNLP}, pages 1538--1548, 2019.

\bibitem[\protect\citeauthoryear{Conneau and Lample}{2019}]{lample2019cross}
A.~Conneau and G.~Lample.
\newblock Cross-lingual language model pretraining.
\newblock In {\em Proc. NeurIPS}, pages 7059--7069, 2019.

\bibitem[\protect\citeauthoryear{der Maaten and
  Hinton}{2008}]{van2008visualizing}
L.~Van der Maaten and G.~Hinton.
\newblock Visualizing data using t-sne.
\newblock {\em JMLR}, 9(11):2579--2605, 2008.

\bibitem[\protect\citeauthoryear{Devlin \bgroup \em et al.\egroup
  }{2019}]{devlin2019bert}
J.~Devlin, M.-W. Chang, K.~Lee, and K.~Toutanova.
\newblock {BERT}: {P}re-training of deep bidirectional transformers for
  language understanding.
\newblock In {\em Proc. NAACL}, pages 4171--4186, 2019.

\bibitem[\protect\citeauthoryear{Ebrahimi and Kann}{2021}]{ebrahimi2021adapt}
A.~Ebrahimi and K.~Kann.
\newblock How to adapt your pretrained multilingual model to 1600 languages.
\newblock In {\em Proc. ACL-IJCNLP}, pages 4555--4567, 2021.

\bibitem[\protect\citeauthoryear{Fukada \bgroup \em et al.\egroup
  }{1992}]{fukada92melcep}
T.~Fukada, K.~Tokuda, T.~Kobayashi, and S.~Imai.
\newblock An adaptive algorithm for mel-cepstral analysis of speech.
\newblock In {\em Proc. ICASSP}, pages 137--140, 1992.

\bibitem[\protect\citeauthoryear{Gouws \bgroup \em et al.\egroup
  }{2015}]{gouws2015bilbowa}
S.~Gouws, Y.~Bengio, and G.~Corrado.
\newblock Bilbowa: Fast bilingual distributed representations without word
  alignments.
\newblock In {\em Proc. ICML}, pages 748--756, 2015.

\bibitem[\protect\citeauthoryear{Gutkin}{2017}]{gutkin2017uniform}
A.~Gutkin.
\newblock Uniform multilingual multi-speaker acoustic model for statistical
  parametric speech synthesis of low-resourced languages.
\newblock In {\em Proc. Interspeech}, pages 2183--2187, 2017.

\bibitem[\protect\citeauthoryear{Hammarstr{\"o}m \bgroup \em et al.\egroup
  }{2021}]{hammarstrom2021glottolog}
H.~Hammarstr{\"o}m, R.~Forkel, M.~Haspelmath, and S.~Bank.
\newblock Glottolog 4.5.
\newblock {\em Max Planck Institute for the Science of Human History}, 2021.

\bibitem[\protect\citeauthoryear{Hayashi \bgroup \em et al.\egroup
  }{2019}]{hayashi2019pre}
T.~Hayashi, S.~Watanabe, T.~Toda, K.~Takeda, S.~Toshniwal, and K.~Livescu.
\newblock Pre-trained text embeddings for enhanced text-to-speech synthesis.
\newblock In {\em Proc. Interspeech}, pages 4430--4434, 2019.

\bibitem[\protect\citeauthoryear{Hayashi \bgroup \em et al.\egroup
  }{2021}]{hayashi2021espnet2}
T.~Hayashi, R.~Yamamoto, T.~Yoshimura, P.~Wu, J.~Shi, T.~Saeki, Y.~Ju,
  Y.~Yasuda, S.~Takamichi, and S.~Watanabe.
\newblock Espnet2-tts: Extending the edge of tts research.
\newblock {\em arXiv preprint arXiv:2110.07840}, 2021.

\bibitem[\protect\citeauthoryear{He \bgroup \em et al.\egroup
  }{2021}]{he2021multilingual}
M.~He, J.~Yang, L.~He, and F.~K Soong.
\newblock Multilingual byte2speech models for scalable low-resource speech
  synthesis.
\newblock {\em arXiv preprint arXiv:2103.03541}, 2021.

\bibitem[\protect\citeauthoryear{Hendrycks and
  Gimpel}{2016}]{hendrycks2016gaussian}
D.~Hendrycks and K.~Gimpel.
\newblock Gaussian error linear units (gelus).
\newblock {\em arXiv preprint arXiv:1606.08415}, 2016.

\bibitem[\protect\citeauthoryear{Jia \bgroup \em et al.\egroup
  }{2021}]{jia2021png}
Y.~Jia, H.~Zen, J.~Shen, Y.~Zhang, and Y.~Wu.
\newblock {PnG BERT}: {A}ugmented {BERT} on phonemes and graphemes for neural
  {TTS}.
\newblock {\em arXiv preprint arXiv:2103.15060}, 2021.

\bibitem[\protect\citeauthoryear{Jr}{2005}]{gordon2005ethnologue}
R.~G~Gordon Jr.
\newblock Ethnologue, languages of the world.
\newblock \url{https://www.ethnologue.com/}, 2005.
\newblock Accessed: 2023-05-27.

\bibitem[\protect\citeauthoryear{Kim \bgroup \em et al.\egroup
  }{2021}]{kim2021conditional}
J.~Kim, J.~Kong, and J.~Son.
\newblock Conditional variational autoencoder with adversarial learning for
  end-to-end text-to-speech.
\newblock In {\em Proc. ICML}, pages 5530--5540, 2021.

\bibitem[\protect\citeauthoryear{Kong \bgroup \em et al.\egroup
  }{2020}]{kong2020hifi}
J.~Kong, J.~Kim, and J.~Bae.
\newblock {HiFi-GAN}: {G}enerative adversarial networks for efficient and high
  fidelity speech synthesis.
\newblock {\em Proc. NeurIPS}, 33:17022--17033, 2020.

\bibitem[\protect\citeauthoryear{Li and Zen}{2016}]{li2016multi}
B.~Li and H.~Zen.
\newblock Multi-language multi-speaker acoustic modeling for {LSTM-RNN} based
  statistical parametric speech synthesis.
\newblock In {\em Proc. Interspeech}, pages 2468--2472, 2016.

\bibitem[\protect\citeauthoryear{Li \bgroup \em et al.\egroup
  }{2019a}]{li2019bytes}
B.~Li, Y.~Zhang, T.~Sainath, Y.~Wu, and W.~Chan.
\newblock Bytes are all you need: End-to-end multilingual speech recognition
  and synthesis with bytes.
\newblock In {\em Proc. ICASSP}, pages 5621--5625, 2019.

\bibitem[\protect\citeauthoryear{Li \bgroup \em et al.\egroup
  }{2019b}]{li2019neural}
N.~Li, S.~Liu, Y.~Liu, S.~Zhao, and M.~Liu.
\newblock Neural speech synthesis with {T}ransformer network.
\newblock In {\em Proc. AAAI}, pages 6706--6713, 2019.

\bibitem[\protect\citeauthoryear{Li \bgroup \em et al.\egroup
  }{2022}]{li2022asr2k}
X.~Li, F.~Metze, D.~R Mortensen, A.~W Black, and S.~Watanabe.
\newblock {ASR2K}: Speech recognition for around 2000 languages without audio.
\newblock {\em arXiv preprint arXiv:2209.02842}, 2022.

\bibitem[\protect\citeauthoryear{Lux and Vu}{2022}]{lux2022language}
F.~Lux and T.~Vu.
\newblock Language-agnostic meta-learning for low-resource text-to-speech with
  articulatory features.
\newblock In {\em Proc. ACL}, pages 6858--6868, 2022.

\bibitem[\protect\citeauthoryear{{Munich Artificial Intelligence Laboratories
  GmbH}}{2017}]{mailabs}
{Munich Artificial Intelligence Laboratories GmbH}.
\newblock The {M-AILABS} speech dataset.
\newblock \url{https://www.caito.de/2019/01/the-m-ailabs-speech-dataset/},
  2017.
\newblock Accessed: 2023-05-27.

\bibitem[\protect\citeauthoryear{Nagrani \bgroup \em et al.\egroup
  }{2017}]{nagrani2017voxceleb}
A.~Nagrani, J.~S. Chung, and A.~Zisserman.
\newblock {VoxCeleb}: A large-scale speaker identification dataset.
\newblock In {\em Proc. Interspeech}, pages 2616--2620, 2017.

\bibitem[\protect\citeauthoryear{Nair and Hinton}{2010}]{nair2010rectified}
V.~Nair and G.~E Hinton.
\newblock Rectified linear units improve restricted boltzmann machines.
\newblock In {\em Proc. ICML}, 2010.

\bibitem[\protect\citeauthoryear{Ni \bgroup \em et al.\egroup
  }{2022}]{ni2022unsupervised}
J.~Ni, L.~Wang, H.~Gao, K.~Qian, Y.~Zhang, S.~Chang, and M.~Hasegawa-Johnson.
\newblock Unsupervised text-to-speech synthesis by unsupervised automatic
  speech recognition.
\newblock In {\em Proc. Interspeech}, pages 461--465, 2022.

\bibitem[\protect\citeauthoryear{Nystrom \bgroup \em et al.\egroup
  }{2015}]{nystrom2015bridges}
N.~A Nystrom, M.~J Levine, R.~Z Roskies, and J~Ray Scott.
\newblock Bridges: a uniquely flexible hpc resource for new communities and
  data analytics.
\newblock In {\em Proc. XSEDE}, pages 1--8, 2015.

\bibitem[\protect\citeauthoryear{Paolacci \bgroup \em et al.\egroup
  }{2010}]{paolacci2010running}
G.~Paolacci, J.~Chandler, and P.~G Ipeirotis.
\newblock Running experiments on amazon mechanical turk.
\newblock {\em Judgment and Decision making}, 5(5):411--419, 2010.

\bibitem[\protect\citeauthoryear{Park and Mulc}{2019}]{park2019css10}
K.~Park and T.~Mulc.
\newblock {CSS10}: A collection of single speaker speech datasets for 10
  languages.
\newblock {\em Proc. Interspeech}, pages 1566--1570, 2019.

\bibitem[\protect\citeauthoryear{Pires \bgroup \em et al.\egroup
  }{2019}]{pires2019multilingual}
T.~Pires, E.~Schlinger, and D.~Garrette.
\newblock How multilingual is multilingual {BERT?}
\newblock In {\em Proc. ACL}, pages 4996--5001, 2019.

\bibitem[\protect\citeauthoryear{Prakash \bgroup \em et al.\egroup
  }{2019}]{prakash2019building}
A.~Prakash, A~L. Thomas, S~Umesh, and H.~A Murthy.
\newblock Building multilingual end-to-end speech synthesisers for {Indian}
  languages.
\newblock In {\em Proc. SSW}, pages 194--199, 2019.

\bibitem[\protect\citeauthoryear{Radford \bgroup \em et al.\egroup
  }{2022}]{radford2022robust}
A.~Radford, J.~W Kim, T.~Xu, G.~Brockman, C.~McLeavey, and I.~Sutskever.
\newblock Robust speech recognition via large-scale weak supervision.
\newblock {\em arXiv preprint arXiv:2212.04356}, 2022.

\bibitem[\protect\citeauthoryear{Ravanelli \bgroup \em et al.\egroup
  }{2021}]{ravanelli2021speechbrain}
M.~Ravanelli, T.~Parcollet, P.~Plantinga, A.~Rouhe, S.~Cornell, L.~Lugosch,
  C.~Subakan, N.~Dawalatabad, A.~Heba, J.~Zhong, et~al.
\newblock {SpeechBrain}: A general-purpose speech toolkit.
\newblock {\em arXiv preprint arXiv:2106.04624}, 2021.

\bibitem[\protect\citeauthoryear{Ren \bgroup \em et al.\egroup
  }{2019}]{ren2019almost}
Y.~Ren, X.~Tan, T.~Qin, S.~Zhao, Z.~Zhao, and T.-Y. Liu.
\newblock Almost unsupervised text to speech and automatic speech recognition.
\newblock In {\em Proc. ICML}, pages 5410--5419, 2019.

\bibitem[\protect\citeauthoryear{Ruder \bgroup \em et al.\egroup
  }{2019}]{ruder2019survey}
S.~Ruder, I.~Vuli{\'c}, and A.~S{\o}gaard.
\newblock A survey of cross-lingual word embedding models.
\newblock {\em JAIR}, 65:569--631, 2019.

\bibitem[\protect\citeauthoryear{Saeki \bgroup \em et al.\egroup
  }{2022a}]{saeki2022utmos}
T.~Saeki, D.~Xin, W.~Nakata, T.~Koriyama, S.~Takamichi, and H.~Saruwatari.
\newblock {UTMOS}: {UTokyo-SaruLab} system for {VoiceMOS} {C}hallenge 2022.
\newblock In {\em Proc. Interspeech}, pages 4521--4525, 2022.

\bibitem[\protect\citeauthoryear{Saeki \bgroup \em et al.\egroup
  }{2022b}]{saeki2022virtuoso}
T.~Saeki, H.~Zen, Z.~Chen, N.~Morioka, G.~Wang, Y.~Zhang, A.~Bapna,
  A.~Rosenberg, and B.~Ramabhadran.
\newblock Virtuoso: Massive multilingual speech-text joint semi-supervised
  learning for text-to-speech.
\newblock {\em arXiv preprint arXiv:2210.15447}, 2022.

\bibitem[\protect\citeauthoryear{Seki \bgroup \em et al.\egroup
  }{2022}]{seki2022text}
K.~Seki, S.~Takamichi, T.~Saeki, and H.~Saruwatari.
\newblock Text-to-speech synthesis from dark data with evaluation-in-the-loop
  data selection.
\newblock {\em arXiv preprint arXiv:2210.14850}, 2022.

\bibitem[\protect\citeauthoryear{Shen \bgroup \em et al.\egroup
  }{2018}]{shen2018natural}
J.~Shen, R.~Pang, R.~J Weiss, M.~Schuster, N.~Jaitly, Z.~Yang, Z.~Chen,
  Y.~Zhang, Y.~Wang, RJ~Skerrv-Ryan, et~al.
\newblock Natural {TTS} synthesis by conditioning {WaveNet} on mel spectrogram
  predictions.
\newblock In {\em Proc. ICASSP}, pages 4779--4783, 2018.

\bibitem[\protect\citeauthoryear{Snyder \bgroup \em et al.\egroup
  }{2018}]{snyder2018x}
D.~Snyder, D.~Garcia-Romero, G.~Sell, D.~Povey, and S.~Khudanpur.
\newblock X-vectors: Robust dnn embeddings for speaker recognition.
\newblock In {\em Proc. ICASSP}, pages 5329--5333, 2018.

\bibitem[\protect\citeauthoryear{Staib \bgroup \em et al.\egroup
  }{2020}]{staib2020phonological}
M.~Staib, T.~H. Teh, A.~Torresquintero, D.~S~R. Mohan, L.~Foglianti, R.~Lenain,
  and J.~Gao.
\newblock Phonological features for 0-shot multilingual speech synthesis.
\newblock In {\em Proc. Interspeech}, pages 2942--2946, 2020.

\bibitem[\protect\citeauthoryear{Tachibana \bgroup \em et al.\egroup
  }{2018}]{tachibana2018efficiently}
H.~Tachibana, K.~Uenoyama, and S.~Aihara.
\newblock Efficiently trainable text-to-speech system based on deep
  convolutional networks with guided attention.
\newblock In {\em Proc. ICASSP}, pages 4784--4788, 2018.

\bibitem[\protect\citeauthoryear{Vaswani \bgroup \em et al.\egroup
  }{2017}]{vaswani2017attention}
A.~Vaswani, N.~Shazeer, N.~Parmar, J.~Uszkoreit, L.~Jones, A.~N Gomez,
  {\L}.~Kaiser, and I.~Polosukhin.
\newblock Attention is all you need.
\newblock In {\em Proc. NeurIPS}, volume~30, 2017.

\bibitem[\protect\citeauthoryear{Veaux \bgroup \em et al.\egroup
  }{2017}]{veaux2017cstr}
C.~Veaux, J.~Yamagishi, K.~MacDonald, et~al.
\newblock {CSTR VCTK} corpus: English multi-speaker corpus for {CSTR} voice
  cloning toolkit.
\newblock {\em University of Edinburgh. The Centre for Speech Technology
  Research (CSTR)}, 2017.

\bibitem[\protect\citeauthoryear{Wang \bgroup \em et al.\egroup
  }{2021}]{wang2021voxpopuli}
C.~Wang, M.~Riviere, A.~Lee, A.~Wu, C.~Talnikar, D.~Haziza, M.~Williamson,
  J.~Pino, and E.~Dupoux.
\newblock {VoxPopuli}: A large-scale multilingual speech corpus for
  representation learning, semi-supervised learning and interpretation.
\newblock In {\em Proc. ACL}, pages 993--1003, 2021.

\bibitem[\protect\citeauthoryear{Watanabe \bgroup \em et al.\egroup
  }{2018}]{watanabe2018espnet}
S.~Watanabe, T.~Hori, S.~Karita, T.~Hayashi, J.~Nishitoba, Y.~Unno, N.-E.~Y.
  Soplin, J.~Heymann, M.~Wiesner, N.~Chen, et~al.
\newblock {ESPnet}: End-to-end speech processing toolkit.
\newblock {\em Proc. Interspeech}, pages 2207--2211, 2018.

\bibitem[\protect\citeauthoryear{Wu and Dredze}{2019}]{wu2019beto}
S.~Wu and M.~Dredze.
\newblock {Beto}, {Bentz}, {Becas}: The surprising cross-lingual effectiveness
  of {BERT}.
\newblock In {\em Proc. EMNLP-IJCNLP}, pages 833--844, 2019.

\bibitem[\protect\citeauthoryear{Xu \bgroup \em et al.\egroup
  }{2021}]{xu2021improving}
G.~Xu, W.~Song, Z.~Zhang, C.~Zhang, X.~He, and B.~Zhou.
\newblock Improving prosody modelling with cross-utterance {BERT} embeddings
  for end-to-end speech synthesis.
\newblock In {\em Proc. ICASSP}, pages 6079--6083, 2021.

\bibitem[\protect\citeauthoryear{Zen \bgroup \em et al.\egroup
  }{2012}]{zen2012statistical}
H.~Zen, N.~Braunschweiler, S.~Buchholz, M.~JF Gales, K.~Knill, S.~Krstulovic,
  and J.~Latorre.
\newblock Statistical parametric speech synthesis based on speaker and language
  factorization.
\newblock {\em TASLP}, 20(6):1713--1724, 2012.

\bibitem[\protect\citeauthoryear{Zen \bgroup \em et al.\egroup
  }{2019}]{zen2019libritts}
H.~Zen, V.~Dang, R.~Clark, Y.~Zhang, R.~J Weiss, Y.~Jia, Z.~Chen, and Y.~Wu.
\newblock {LibriTTS}: {A} corpus derived from {LibriSpeech} for text-to-speech.
\newblock In {\em Proc. Interspeech}, pages 1526--1530, 2019.

\bibitem[\protect\citeauthoryear{Zhang and Lin}{2020}]{zhang2020unsupervised}
H.~Zhang and Y.~Lin.
\newblock Unsupervised learning for sequence-to-sequence text-to-speech for
  low-resource languages.
\newblock {\em Proc. Interspeech}, pages 3161--3165, 2020.

\bibitem[\protect\citeauthoryear{Zhang \bgroup \em et al.\egroup
  }{2022}]{zhang2022mixed}
G.~Zhang, K.~Song, X.~Tan, D.~Tan, Y.~Yan, Y.~Liu, G.~Wang, W.~Zhou, T.~Qin,
  T.~Lee, et~al.
\newblock {Mixed-Phoneme BERT}: {I}mproving {BERT} with mixed phoneme and
  sup-phoneme representations for text to speech.
\newblock In {\em Proc. Interspeech}, pages 456--460, 2022.

\end{thebibliography}

\newpage

\appendix
\section{Text data for pretraining}\label{sec:eval-text-data}

We investigated the effect of different types of text data used in pretraining $\mathcal{D}_{\mathrm{text}}$ on the performance of our method.
As described in \S~\ref{sec:eval-setting}, we used spoken texts from VoxPopuli, M-AILABS, and CSS10 in the evaluations presented in \S~\ref{sec:eval}.
However, we can acquire a large amount of text data from datasets designed for NLP or from web-crawled text resources. These data typically consist of written texts and cover a wide range of domains.
To investigate the effectiveness of using written texts for our multilingual TTS, we used ParaCrawl~\cite{banon2020paracrawl}, a web-crawled text dataset built for machine translation, for $\mathcal{D}_{\mathrm{text}}$ during the unsupervised pretraining described in \S~\ref{sec:method-pretrain}.
For the investigation, we randomly sampled the same amount of texts as in Table~\ref{tab:data} for each language.
We then trained our model using the following three different cases. 1) \textit{Spoken Text}: Only using the spoken text for pretraining as in the previous evaluations, 2) \textit{Written Text}: Only using the text data from ParaCrawl, and 3) \textit{Spoken+Written Text}: We combined the text data in \textit{Spoken Text} and \textit{Written Text}.
We used the byte-based proposed model presented in \S~\ref{sec:eval-seen} and \S~\ref{sec:eval-unseen}.
Table~\ref{tab:text-domain} lists the results.

We observed that \textit{Spoken Text} outperformed \textit{Written Text} in all the metrics and languages, resulting in an average difference of 0.3 in MCD and 2.94\% in CER. These results demonstrate the effectiveness of using spoken text for pretraining.
\textit{Spoken+Written Text} showed on average 0.11 lower MCD and 0.46\% higher CER compared to \textit{Spoken Text}.
However, for the unseen language, \textit{Spoken+Written Text} outperformed \textit{Spoken Text} in MCD and CER.
These results suggest that adding written text data can improve the generalization of our TTS models for the zero-shot scenarios.

\begin{table*}[tp]
\centering
\scalebox{0.95}{
\begin{tabular}{l|cccccccc|cc|cc}
\toprule
\multirow{3}{*}{Method} & \multicolumn{8}{c|}{Seen}                                                                                                         & \multicolumn{2}{c|}{Unseen} & \multicolumn{2}{c}{\multirow{2}{*}{Avg.}} \\
                        & \multicolumn{2}{c|}{de}          & \multicolumn{2}{c|}{fr}          & \multicolumn{2}{c|}{ru}           & \multicolumn{2}{c|}{fi} & \multicolumn{2}{c|}{es}     & \multicolumn{2}{c}{}                     \\
                        & MCD  & \multicolumn{1}{c|}{CER}  & MCD  & \multicolumn{1}{c|}{CER}  & MCD  & \multicolumn{1}{c|}{CER}   & MCD        & CER        & MCD          & CER          & MCD                & CER                 \\ \midrule
Spoken Text             & 5.65 & \multicolumn{1}{c|}{3.79} & 6.48 & \multicolumn{1}{c|}{7.15} & 7.38 & \multicolumn{1}{c|}{10.62} & 4.99       & 5.28       & 9.05         & 18.27        & 6.46               & \textbf{9.58}                \\
Written Text           & 5.81 & \multicolumn{1}{c|}{4.55} & 6.94 & \multicolumn{1}{c|}{9.10} & 7.61 & \multicolumn{1}{c|}{21.24} & 5.22       & 12.73      & 9.50         & 18.44        & 6.76               & 12.52               \\
Spoken+Written Text    & 5.54 & \multicolumn{1}{c|}{3.72} & 6.34 & \multicolumn{1}{c|}{7.51} & 7.07 & \multicolumn{1}{c|}{15.33} & 4.96       & 5.44       & 8.82         & 17.48        & \textbf{6.35}               & 10.04               \\ \bottomrule
\end{tabular}
}
\caption{Comparison of text data domain for unsupervised text pretraining.}
\label{tab:text-domain}
\end{table*}

\begin{table*}[tp]
\centering
\scalebox{0.95}{
\begin{tabular}{l|cccccccc|cc|cc}
\toprule
\multirow{3}{*}{Method} & \multicolumn{8}{c|}{Seen}                                                                                                         & \multicolumn{2}{c|}{Unseen} & \multicolumn{2}{c}{\multirow{2}{*}{Avg.}} \\
                        & \multicolumn{2}{c|}{de}          & \multicolumn{2}{c|}{fr}          & \multicolumn{2}{c|}{ru}           & \multicolumn{2}{c|}{fi} & \multicolumn{2}{c|}{es}     & \multicolumn{2}{c}{}                     \\
                        & MCD  & \multicolumn{1}{c|}{CER}  & MCD  & \multicolumn{1}{c|}{CER}  & MCD  & \multicolumn{1}{c|}{CER}   & MCD        & CER        & MCD          & CER          & MCD                & CER                 \\ \midrule
Residual layer       & 5.65 & \multicolumn{1}{c|}{3.79} & 6.48 & \multicolumn{1}{c|}{7.15} & 7.38 & \multicolumn{1}{c|}{10.62} & 4.99       & 5.28       & 9.05         & 18.27        & 6.46               & \textbf{9.58}             \\
Transformer encoder           & 5.77 & \multicolumn{1}{c|}{4.63} & 6.36 & \multicolumn{1}{c|}{6.61} & 7.17 & \multicolumn{1}{c|}{11.25} & 4.90       & 6.50    & 8.89         & 14.15        & \textbf{6.44}               & 9.97   \\ \bottomrule
\end{tabular}
}
\caption{Comparison of bottleneck layer architecture.}
\label{tab:arch-bottle}
\end{table*}

\begin{table*}[tp]
\centering
\scalebox{0.92}{
\begin{tabular}{l|cccccc|cccccc}
\toprule
\multirow{2}{*}{Method}              & \multicolumn{2}{c}{ru}            & \multicolumn{2}{c}{hu}           & \multicolumn{2}{c|}{fi} & \multicolumn{2}{c}{de}             & \multicolumn{2}{c|}{fr}             & \multicolumn{2}{c}{es} \\
                                     & MCD  & \multicolumn{1}{c|}{CER}   & MCD  & \multicolumn{1}{c|}{CER}  & MCD        & CER        & MCD   & \multicolumn{1}{c|}{CER}   & MCD   & \multicolumn{1}{c|}{CER}   & MCD        & CER       \\ \midrule
Original (de, fr, nl, fi, hu, ru, el) & 7.38 & \multicolumn{1}{c|}{10.62} & 5.01 & \multicolumn{1}{c|}{6.05} & 4.99       & 5.28       & 5.65  & \multicolumn{1}{c|}{3.79}  & 6.48  & \multicolumn{1}{c|}{7.15}  & 9.05       & 18.27     \\
Excluded (fi, hu, ru, el)             & 7.00 & \multicolumn{1}{c|}{11.11} & 5.32 & \multicolumn{1}{c|}{6.92} & 4.98       & 5.46       & 10.39 & \multicolumn{1}{c|}{34.11} & 10.90 & \multicolumn{1}{c|}{49.65} & 10.00      & 24.80     \\ \bottomrule
\end{tabular}
}
\caption{Effects of excluding languages from paired data.}
\label{tab:exclude-langs}
\end{table*}

\section{Architecture of bottleneck layer}\label{sec:eval-arch-bottle}
As described in \S~\ref{sec:method-architecture}, we used the residual layer for our bottleneck layer.
Also, we demonstrated the effectiveness of the residula bottleneck layer for both seen and unseen languages in the evaluation presented in \S~\ref{sec:eval-ablation}.
In this section, we explored an alternative architecture for the bottleneck layer.
We conducted experiments using a single-layer Transformer encoder as the bottleneck layer (referred to as \textit{Transformer-encoder}), comparing it with the original residual layer detailed in \S~\ref{sec:method-architecture} (referred to as \textit{Residual layer}).
Table~\ref{tab:arch-bottle} lists the results.

For the seen languages, the superior performance between the residual layer and the transformer encoder varied, depending on the specific language and evaluation metrics. However, for the unseen language, the Transformer encoder showed higher performance, achieving an improvement of 4.12 in CER. Looking at the average scores across all languages, the Transformer encoder had a slightly lower MCD, while the CER was reduced by 0.39 when using the residual layer. These results suggest that the use of a deeper layer can improve the generalizability of the proposed model. Nevertheless, the overall performance of both models remains comparable in terms of average metrics.

\section{Effect of excluding some languages from paired data.}\label{sec:eval-exclusion}

In this section, we have deliberately excluded several languages from the paired data used for the supervised learning described in \S~\ref{sec:method-supervised} in order to study their impact.
As shown in Table~\ref{tab:data}, the paired data originally included the languages de, fr, nl, fi, hu, ru, and el.
As a comparison case, we first excluded fr, which belongs to Italic languages as es according to Glottolog~\cite{hammarstrom2021glottolog}.
We also removed de and nl, which belong to Germanic languages, from the paired data.
Consequently, in the comparison case, supervised learning was performed only with fi, ru, hu, and el.
Table~\ref{tab:exclude-langs} lists the results.
\textit{Original} corresponds to the case shown in Table~\ref{tab:data}, while \textit{Excluded} denotes the case where only fi, ru, hu, and el were used.

The three languages listed on the left-hand side of Table~\ref{tab:exclude-langs} (ru, hu, fi) represent the \textit{seen} languages in both cases.
Interestingly, the \textit{Original} scenario generally outperformed the \textit{Excluded} scenario for these languages. These results indicate that in the context of multilingual TTS training, performance can potentially be improved by including a wider variety of languages rather than restricting to similar languages.
This suggests the effectiveness of massively multilingual TTS training, as supported by previous work~\cite{saeki2022virtuoso}.

The two languages (de, fr) were included in the paired data for \textit{Original}, but were excluded from the paired data for \textit{Excluded}.
We observed that the zero-shot setting resulted in a decrease in performance.
Comparing the de results in Table~\ref{tab:diff-unseen} with those in Table~\ref{tab:exclude-langs}, we confirmed that excluding fr, nl, and es resulted in a 6.1\% performance degradation in CER.
Also, es was absent from the paired data in both the \textit{Original} and \textit{Excluded} scenarios.
An examination of the es results revealed a 6.53\% increase in CER for the \textit{Excluded} scenario.
These results demonstrate the importance of including linguistically similar languages in the paired data.

\begin{figure}
    \centering
    \includegraphics[width=1.0\linewidth, clip]{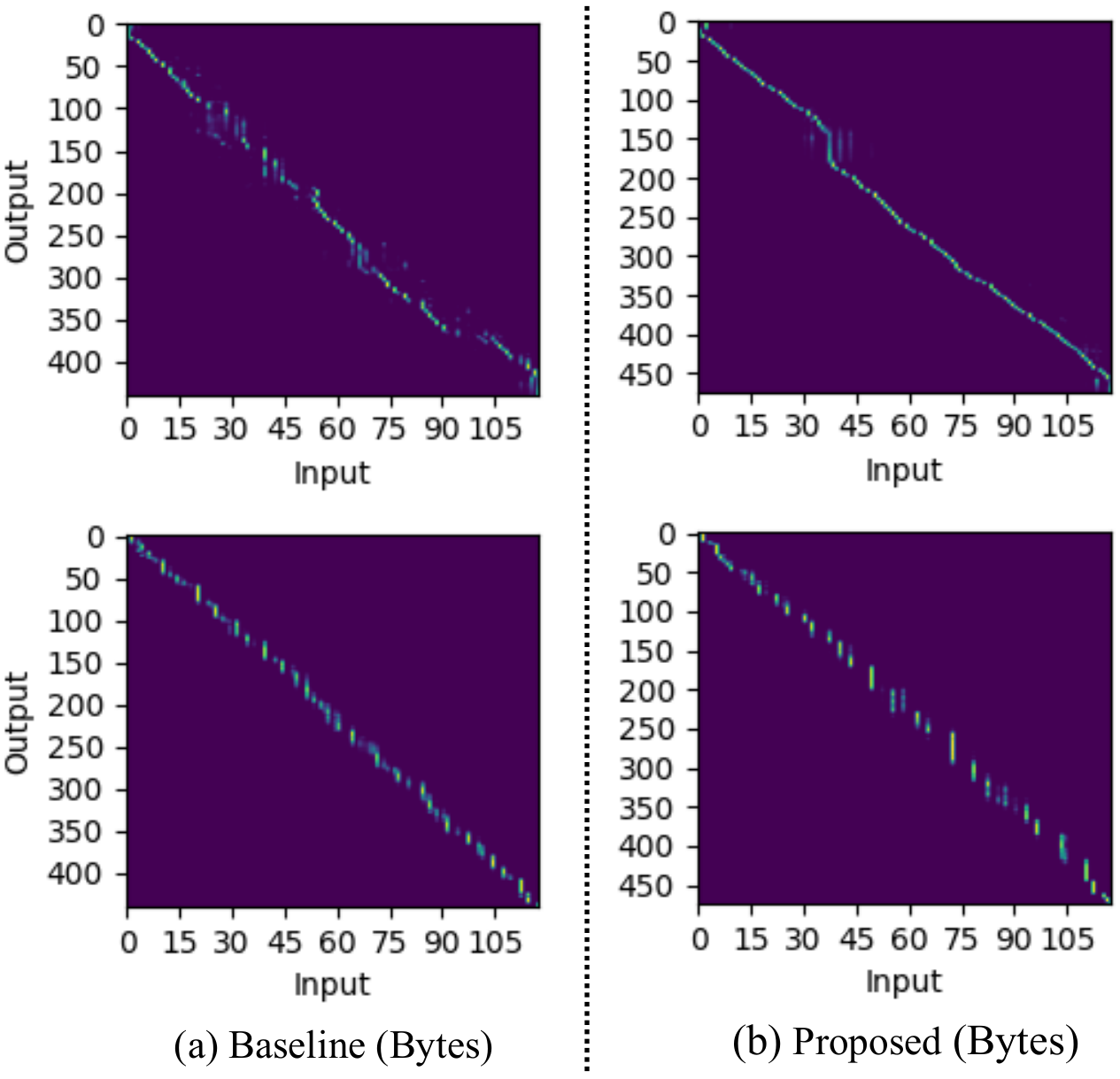}
    \caption{Cross-attention maps obtained from byte-based baseline and proposed methods, which correspond to first attention head in fifth and sixth layers of decoder.}
    \label{fig:cross-attention}
\end{figure}

\section{Observation of cross-attention map.}\label{sec:eval-align}

In this section, we show the cross-attention maps obtained during inference from both the byte-based proposed and the baseline methods defined in \S~\ref{sec:eval-setting-baselines}.
In both cases, an utterance sampled from an unseen language (Spanish) was used.
Fig.~\ref{fig:cross-attention} shows the cross-attention maps corresponding to the first attention head in the fifth and sixth layers of the decoder.
It should be noted that we have cross-attention maps for other heads and layers.
The results shown in Fig.~\ref{fig:cross-attention} are derived from attention maps that have a diagonal shape in higher layers for both the baseline and the proposed methods.
The top part of Fig.~\ref{fig:cross-attention} represents results from the fifth layer, while the bottom part corresponds to results from the sixth layer.
The left side of the figure shows the results of the baseline method, while the right side shows the results of the proposed method.

We observe that the fifth layer of the baseline model shows a discontinuity in the attention map, which leads to instability of the linguistic content.
Conversely, in the fifth layer of the proposed model, the attention map is significantly more continuous than in the baseline method.
These results suggest that our unsupervised text pretraining can improve cross-attention in the absence of paired speech-text data.
The results are also reflected in the intelligibility difference between the baseline and the proposed methods presented in \S~\ref{sec:eval-unseen}.

\end{document}